\documentclass[showpacs,floatfix,citeautoscript,twocolumn,prb,superscriptaddress]{revtex4-1}

\usepackage{graphicx,xcolor}
\usepackage{textgreek}
\usepackage{upgreek}
\usepackage{siunitx}
\usepackage{amsmath}
\usepackage{comment}
\usepackage{textcomp}
\usepackage{inputenc}
\usepackage{caption}
\usepackage{subcaption}
\usepackage{multirow}
\usepackage{adjustbox}
\usepackage{gensymb}
\usepackage{xr}
\usepackage{hyperref}
\usepackage{cleveref}

\newcommand*{\tsi}{(TaSe$_4$)$_2$I}

\makeatletter

\newcommand*{\addFileDependency}[1]{
\typeout{(#1)}
\@addtofilelist{#1}
\IfFileExists{#1}{}{\typeout{No file #1.}}
}\makeatother

\newcommand*{\myexternaldocument}[1]{%
\externaldocument{#1}%
\addFileDependency{#1.tex}%
\addFileDependency{#1.aux}%
}

\myexternaldocument{supplementary}

\begin{document}

\title{Disorder and diffuse scattering in single-chirality (TaSe$_4$)$_2$I crystals}

\author{Jacob A. Christensen}
\affiliation{Materials Research Laboratory, University of Illinois at Urbana-Champaign, Urbana, IL 61801, USA}
\affiliation{Materials Science and Engineering Department, University of Illinois at Urbana-Champaign, Urbana, IL 61801, USA}
\author{Simon Bettler}
\author{Kejian Qu}
\affiliation{Department of Physics, University of Illinois at Urbana-Champaign, Urbana, IL 61801, USA}
\affiliation{Materials Research Laboratory, University of Illinois at Urbana-Champaign, Urbana, IL 61801, USA}
\author{Jeffrey Huang}
\affiliation{Materials Research Laboratory, University of Illinois at Urbana-Champaign, Urbana, IL 61801, USA}
\affiliation{Materials Science and Engineering Department, University of Illinois at Urbana-Champaign, Urbana, IL 61801, USA}
\author{Soyeun Kim}
\author{Yinchuan Lu}
\affiliation{Department of Physics, University of Illinois at Urbana-Champaign, Urbana, IL 61801, USA}
\affiliation{Materials Research Laboratory, University of Illinois at Urbana-Champaign, Urbana, IL 61801, USA}
\author{Chengxi Zhao}
\affiliation{Materials Research Laboratory, University of Illinois at Urbana-Champaign, Urbana, IL 61801, USA}
\affiliation{Materials Science and Engineering Department, University of Illinois at Urbana-Champaign, Urbana, IL 61801, USA}
\author{Jin Chen}
\affiliation{Department of Physics, University of Illinois at Urbana-Champaign, Urbana, IL 61801, USA}
\affiliation{Materials Research Laboratory, University of Illinois at Urbana-Champaign, Urbana, IL 61801, USA}
\author{Matthew J. Krogstad}
\affiliation{Advanced Photon Source, Argonne National Laboratory, Lemont IL, 60439, USA}
\author{Toby J. Woods}
\affiliation{Department of Chemistry, University of Illinois at Urbana-Champaign, Urbana, IL 61801, USA}
\author{Fahad Mahmood}
\affiliation{Department of Physics, University of Illinois at Urbana-Champaign, Urbana, IL 61801, USA}
\affiliation{Materials Research Laboratory, University of Illinois at Urbana-Champaign, Urbana, IL 61801, USA}
\author{Pinshane Y. Huang}
\affiliation{Materials Research Laboratory, University of Illinois at Urbana-Champaign, Urbana, IL 61801, USA}
\affiliation{Materials Science and Engineering Department, University of Illinois at Urbana-Champaign, Urbana, IL 61801, USA}
\author{Peter Abbamonte}
\affiliation{Department of Physics, University of Illinois at Urbana-Champaign, Urbana, IL 61801, USA}
\affiliation{Materials Research Laboratory, University of Illinois at Urbana-Champaign, Urbana, IL 61801, USA}
\author{Daniel P. Shoemaker}\email{dpshoema@illinois.edu}
\affiliation{Materials Research Laboratory, University of Illinois at Urbana-Champaign, Urbana, IL 61801, USA}
\affiliation{Materials Science and Engineering Department, University of Illinois at Urbana-Champaign, Urbana, IL 61801, USA}


\begin{abstract}
The quasi-one-dimensional chiral compound (TaSe$_4$)$_2$I has been extensively studied as a prime example of a topological Weyl semimetal. Upon crossing its phase transition temperature $T$\textsubscript{CDW} $\approx 263$~K, (TaSe$_4$)$_2$I exhibits incommensurate charge density wave (CDW) modulations described by the well-defined propagation vector $\sim$(0.05, 0.05, 0.11), oblique to the TaSe$_4$ chains.
Although optical and transport properties greatly depend on chirality, there is no systematic report about chiral domain size for (TaSe$_4$)$_2$I. In this study, our single-crystal scattering refinements reveal a bulk iodine deficiency, and Flack parameter measurements on multiple crystals demonstrate that separate (TaSe$_4$)$_2$I crystals have uniform handedness, supported by direct imaging and helicity dependent THz emission spectroscopy.
Our single-crystal X-ray scattering and calculated diffraction patterns identify multiple diffuse features and create a real-space picture of the temperature-dependent (TaSe$_4$)$_2$I crystal structure.
The short-range diffuse features are present at room temperature and decrease in intensity as the CDW modulation develops.
These transverse displacements, along with electron pinning from the iodine deficiency, help explain why \tsi\ behaves as an electronic semiconductor at temperatures above and below $T$\textsubscript{CDW}, despite a metallic band structure calculated from density functional theory of the ideal structure.





\end{abstract}

\maketitle 

\section{Introduction} 

Materials with nontrivial topological band structures represent an exciting frontier for exotic physics and quantum materials applications \cite{wieder_topological_2022}. Topological insulators show strong spin-orbit coupling leading to inverted and gapped electronic bands, resulting in topologically protected metallic surface states and insulated bulk states \cite{fu_topological_2007,moore_topological_2007,qi_topological_2011,kumar_topological_2021, hasan_colloquium_2010, moore_birth_2010, b_a_bernevig_topological_2013}. Some electronic structures with similar band inversion may still retain isolated band crossings; such materials are known as Weyl or Dirac semimetals \cite{wan_topological_2011,yan_topological_2017, narang_topology_2021, kumar_topological_2021}. Much attention has recently fallen on the quasi-one-dimensional chiral charge density wave (CDW) material (TaSe$_4$)$_2$I, a prototype Weyl semimetal. \tsi\ belongs to a class of materials characterized by exotic Fermi arc topological surface states and Weyl fermions within the bulk \cite{shi_charge-density-wave_2021,weyl_elektron_1929,wan_topological_2011, potter_quantum_2014,xu_discovery_2015, jia_weyl_2016, hasan_discovery_2017}. Weyl fermions are massless quasi-particles that manifest themselves as low-energy excitations of the Weyl semimetal; when at zero energy, these quasi-particles correspond to degenerate band crossings found near the Fermi surface where the momentum dispersion is approximately linear in all three dimensions, serving as a 3D analog to 2D materials like graphene \cite{castro_neto_electronic_2009}. Known as Weyl nodes, these band crossings occur as pairs with opposite chirality, requiring either time-reversal symmetry or inversion symmetry to be broken \cite{weng_weyl_2015}. Weyl node pairs, with their opposite topological charges, act as sources (monopoles) and sinks (antimonopoles) for Berry curvature \cite{b_a_bernevig_topological_2013}, and their separation in momentum space provides a topologically protected phase \cite{liu_weyl_2014,hasan_discovery_2017}.

\begin{figure}
\centering\includegraphics[width=\columnwidth]{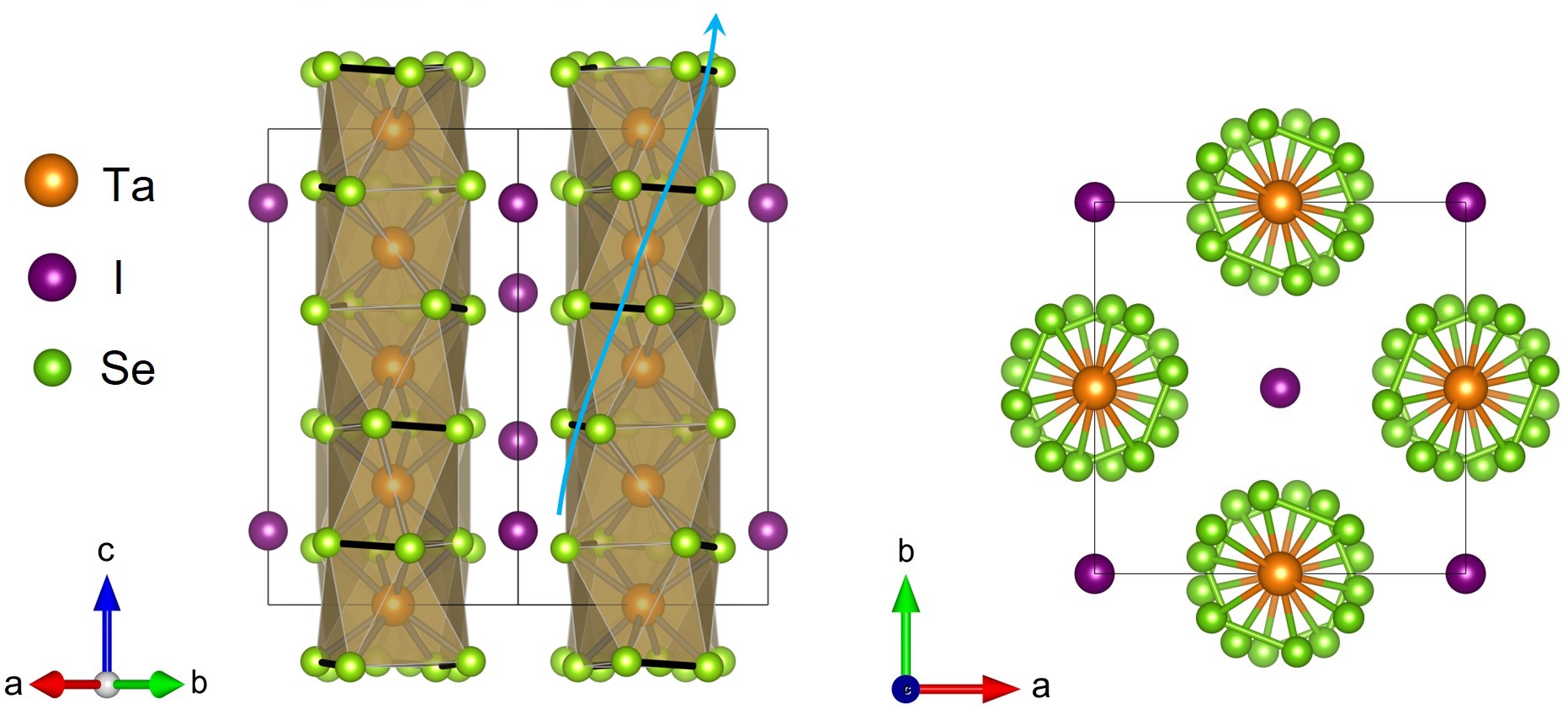} \\
\caption{\label{fig:TSI_unit_cell}
The room temperature $I422$ \tsi\ unit cell crystal structure, shown from a side and top ([001]) perspective. Chains of Ta form along the $c$ direction, each coordinated by two rectangles of Se with alternating Se--Se bonds shown in black. The orientation of the Se--Se bond down the chain imparts a handedness denoted by the blue arrow.  Iodine ions separate each chiral chain.
} 
\end{figure}

The structural properties of \tsi\ were first published decades ago \cite{gressier_preparation_1982, gressier_electronic_1984}. At high temperature, chiral \tsi\ crystallizes in the tetragonal space group $I422$ with each body-centered unit cell consisting of two parallel TaSe$_4$ chains, where each Ta ion is 8-coordinated by a pair of Se$_4$ rectangles, while each rectangle contains a pair of $Se_2^{2-}$ dimers. These equally-spaced chains are separated by a row of four iodine ions as shown in Fig. \ref{fig:TSI_unit_cell}. The orientation of Se imparts a chiral handedness. Topological semimetals lacking mirror symmetry in this manner are theorized to display many unusual optical, magnetic, and transport properties \cite{jia_weyl_2016, zhong_gyrotropic_2016, sanchez_topological_2019, schroter_chiral_2019} absent in centrosymmetric semimetals,
such as helicity dependent currents produced through the circular photogalvanic effect (CPGE)\cite{asnin_circular_1979, belinicher_photogalvanic_1980, de_juan_quantized_2017, rees_helicity-dependent_2020}, a phenomenon leveraged in this paper.

\tsi\ enters an incommensurate CDW state at the transition temperature $T$\textsubscript{CDW} $\approx 263$~K, commonly detected by an anomaly in electrical resistivity.\cite{maki_charge_1983,wang_charge_1983} Weyl semimetals may be gapped out when Weyl nodes are spontaneously removed through pair annihilation between fermions and holes, creating a less-explored topological configuration known as an axionic insulator in tandem with a CDW \cite{wang_chiral_2013, sekine_axion_2021}. Although evidence has been proposed for this behavior in \tsi\ \cite{gooth_axionic_2019}, it is not entirely clear if the CDW phase is a direct result of chiral symmetry breaking in this manner\cite{sinchenko_does_2022}. In general, the metallic state in one-dimensional systems similar to \tsi\ is unstable, prone to breaking symmetry by formation of spin or charge density wave phases \cite{Gruner_2019}. 
The \tsi\ CDW phase is often described as Peierls-like in nature \cite{voit_electronic_2000}. According to this picture, distortions are induced by electronic instabilities originating from Fermi surface nesting found in quasi-1D systems \cite{monceau_electronic_2012,tournier-colletta_electronic_2013}. 
Previous \tsi\ band structure studies also lend to this \cite{gressier_electronic_1984, zhang_first-principles_2020}.
But detailed X-ray and neutron scattering analyses have revealed a more complicated modulation than a textbook one-dimensional Peierls instability: broken symmetry is indeed found along the chain direction as a Ta-tetramerization, but strong acoustic modulations form perpendicular to both the chain direction and the modulation wavevectors \cite{lee_x-ray_1985,lorenzo_neutron_1998,smaalen_structure_2001,favre-nicolin_structural_2001}. 
Ascertaining the true structural and electronic behavior of the CDW phase is challenging due to incommensurability, but calculation of the band structure of a model \tsi\ cell with Ta displacements by Tournier-Colletta, et al.\cite{tournier-colletta_electronic_2013} confirmed the creation of a gap near $q_\textrm{CDW}/2$, which they correlate with angle-resolved photoemission spectroscopy (ARPES) data. This was only done at 100~K, so the possible absence of a gap at high temperature was not examined. The semiconductor-like transport at high temperatures was conjectured to occur due to polaron hopping from the iodine vacancies \cite{tournier-colletta_electronic_2013}.

In the above studies, substantial work on structural characterization and transport has been performed, while the ultimate goal of helicity-dependent transport and optical excitation requires operation on a domain of \tsi\ with single chirality (one enantiomer).
The challenge of determining enantiopurity in a dense, opaque material can be addressed by multiple methods.
Here we obtain both chirality and occupancy information from an extensive set of single-crystal X-ray refinements which reveal a consistent, intrinsic iodine deficiency within bulk (TaSe$_4$)$_2$I. The handedness of the refinements suggest that each \tsi\ crystal tends to form with uniform handedness, which is affirmed by annular dark field scanning transmission electron microscopy (ADF-STEM) images and measurements of the circular photogalvanic effect. Synchrotron single-crystal X-ray diffraction (XRD) measurements above and below the CDW transition reveal organized diffuse scattering, often with a strong component in the $ab$-plane, confirming the strong effects between chemical disorder and the CDW transition in \tsi. The source and nature of \tsi\ diffuse scattering is investigated through scattering calculated from discrete atomic models.

\section{Methods}

Single crystals of \tsi\ were grown with a recipe similar to what was described by Maki, et al. \cite{maki_charge_1983} Stoichiometric amounts of iodine granules (Fisher, 99.99\%), selenium powder (Alfa Aesar, 99.999\%), and tantalum wire (Fisher, 99.95\%) were loaded and sealed in 15 mm diameter evacuated quartz tubes along with an excess of iodine to balance the loss of iodine during sample handling. A thermal gradient of 600 $^{\circ}$C to 500 $^{\circ}$C was applied for 10 days, producing black, thin crystals up to 12 mm in length and up to 2 mm in width, with typical dimensions being $3 \times 1 \times 1$ mm\textsuperscript{3}. Smaller reaction tubes and tantalum powder were also used, but the crystals were not as large as those grown in the way reported above. Crystal images are shown in Fig. \ref{fig:Ta2Se8I_optical_image}. A powder XRD fit is provided in the Supplemental Material (Fig. S3)\cite{supplement}.


For our diffuse scattering investigations, single-crystal reciprocal space scattering maps were obtained at the Advanced Photon Source (APS) beamline 6ID-D using a beam energy of 64.8 keV in a transmission geometry.  The \tsi\ sample with a diameter of about 0.3 mm was mounted on the tip of a Kapton capillary. A Pilatus 2M CdTe detector placed at a distance of 125.6 cm from the sample was used to collect energy-integrated images with higher dynamic range than our in-lab source described in the following paragraph. Sample temperatures were controlled between 30 and 300 K using a Oxford N-HeliX Cryostream. The raw images were pre-processed with a peak-finding algorithm to determine and refine the orientation matrix and then rebinned with a bin size of $0.02\times0.02\times0.02$ rlu into a reciprocal space volume $\pm$ $10(\textrm{H})\times10(\textrm{K})\times12(\textrm{L})$ rlu using the CCTW reduction workflow \cite{krogstad_reciprocal_2020,Jennings_2016}.


To track the chirality trends of the material, additional room temperature data were collected on a Bruker D8 Venture diffractometer equipped with a Photon-II CPAD detector and Kappa goniometer. An Iµs microfocus Mo source ($\lambda$ = 0.71073 \AA) coupled with a multi-layer mirror monochromator provided the incident beam. The sample was mounted on a 0.3 mm nylon loop with a minimal amount of Paratone-N oil. A single 180\degree{} $\phi$-scan was collected for each crystal fragment with 1\degree{} frame angles. The collection, cell refinement, and integration of intensity data were carried out with the APEX3 software and SHELXL \cite{sheldrick_crystal_2015}. Multi-scan absorption corrections were performed with SADABS2.

Specimens of \tsi\ were prepared for ADF-STEM using focused ion beam lift-out and milling (micrographs provided in Fig. S4 of the Supplemental Material \cite{supplement}). Imaging was performed using an aberration-corrected STEM (Thermo Fisher Scientific Themis Z) operated at 300 kV, with 18 mrad convergence semiangle and 30 pA probe current.  Individual frames were acquired with dwell time of 1 $\mu$s and pixel size of 8.7 pm, and 10-50 frames were averaged to produce the final images. 

THz emission spectroscopy was performed using a custom built time-domain THz spectroscopy setup based on a
Yb:KGW amplifier laser (PHAROS, Light Conversion). The fundamental laser pulse wavelength is 1030 nm with a
pulse duration of $\sim$160 fs. The fundamental beam is split into pump and probe paths using a 90:10 beamsplitter.
The pump had a 1/e\textsuperscript{2} width of $\sim$1 mm on the sample at a 45° angle of incidence. The THz field $E_\textrm{THz}(t)$ radiated by the sample is collected by an off-axis parabolic mirror and focused on a CdTe (110) crystal. The electro-optical sampling probe beam is made to spatially and temporally overlap with $E_\textrm{THz}(t)$ on the CdTe crystal for electro-optic sampling. The path length of the probe is adjusted with a delay stage to control the time delay $(t_\textrm{delay})$.

All lattice models and diffuse scattering were simulated using the DISCUS software suite \cite{proffen_discus_1997,proffen_discus_1999,Neder_Proffen_2008}, applying both periodic displacement and density modulations. The scattering was simulated on $10\times10\times10$ lattices built with the standard \tsi\ unit cell displayed in Fig. \ref{fig:TSI_unit_cell}. The extent and resolution of the simulated reciprocal space was adjusted to avoid finite size effects (see Supplemental Material \cite{supplement} for more details). For most of our results the reciprocal volume was set to $10(\textrm{H})\times10(\textrm{K})\times10(\textrm{L})$ rlu with 200 points along each axis. All diffuse scattering simulations were calculated using MoK$\alpha$ X-rays ($\lambda$ = 0.71073 \AA) and with both atomic displacement parameters and anomalous dispersion turned off.

\begin{figure}
\centering\includegraphics[width=\columnwidth]{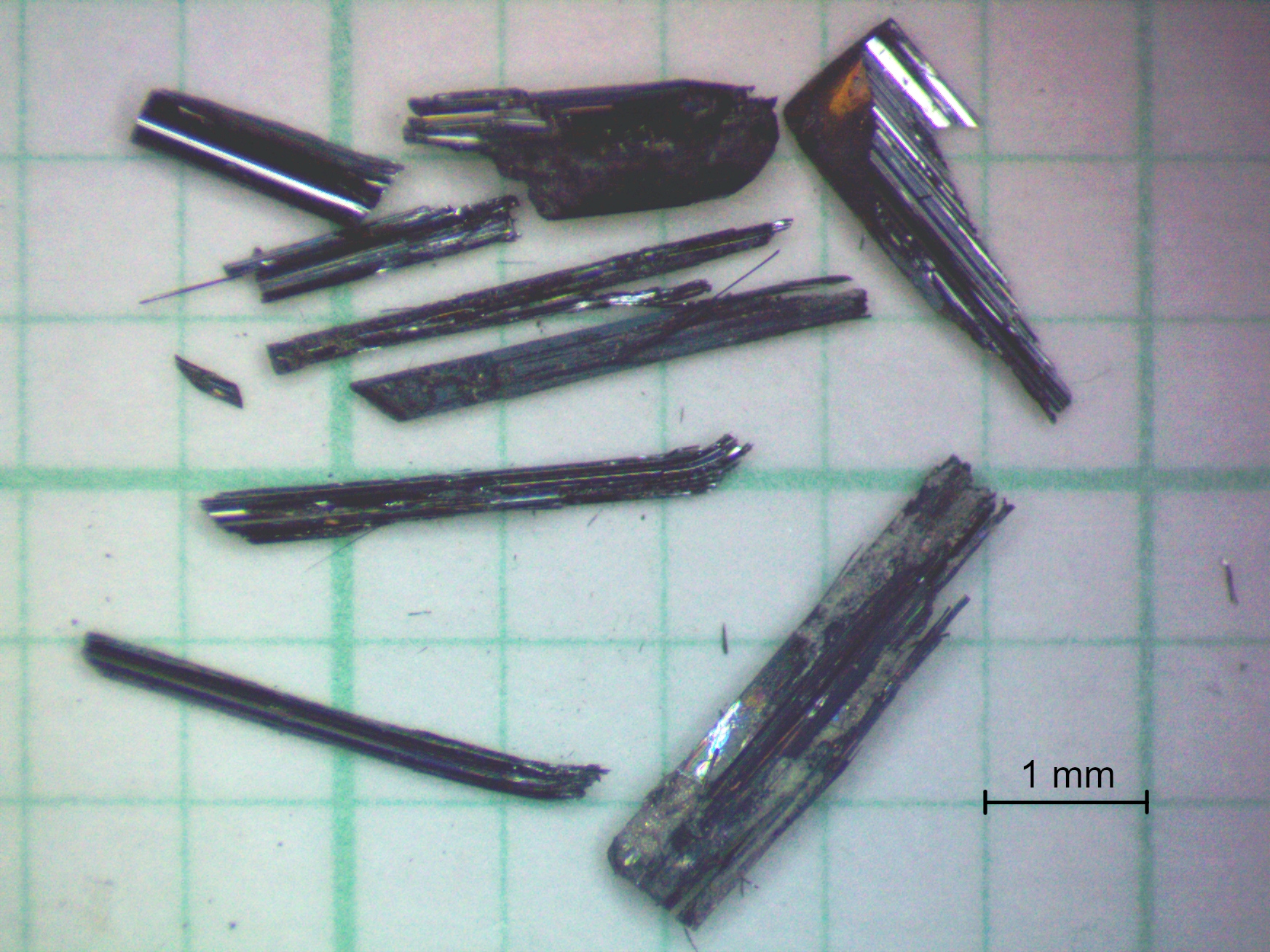} \\
\caption{\label{fig:Ta2Se8I_optical_image}
Examples of the \tsi\ crystals retrieved from the CVT reaction. Large crystals are brittle and composed of thin, needle-like slivers which break apart when cut.
} 
\end{figure}

\section{Results and Discussion}

\subsection{Single-crystal structural refinement} \label{refinement}

\begin{table}[hbt!]
       \caption{\label{table:full_refinement_table}
    Parameters from four full data set single-crystal X-ray solutions. Samples 1 and 2 are separate crystals at room temperature, while sample 3 and 4 are the same crystal measured at room temperature and $T = 100$ K, respectively. A Goodness of Fit (GooF) value of 1 indicates a high-quality refinement. A Flack parameter value near zero indicates a proper configuration with no inversion. Each crystal's iodine deficiency is given in terms of the fraction of full occupation (final column).} \begin{adjustbox}{width=\columnwidth,center}
        \centering
        \small
        \begin{tabular}{ccccc}
        \hline
            \textbf{Sample} & \textbf{Handedness} & \textbf{GooF} & \textbf{Flack} & \textbf{Occupancy} \\ \hline
            1 & left & 1.149 & 0.017(26) & 0.886(22) \\ 
            2 & left & 1.071 & -0.012(74) & 0.888(50) \\ 
            3 & left & 1.073 & 0.015(24) & 0.882(24) \\ 
            4 & left & 1.235 & 0.021(15) & 0.873(53) \\ \hline
        \end{tabular}
    \end{adjustbox}

\end{table}


\begin{figure}
\centering\includegraphics[width=\columnwidth]{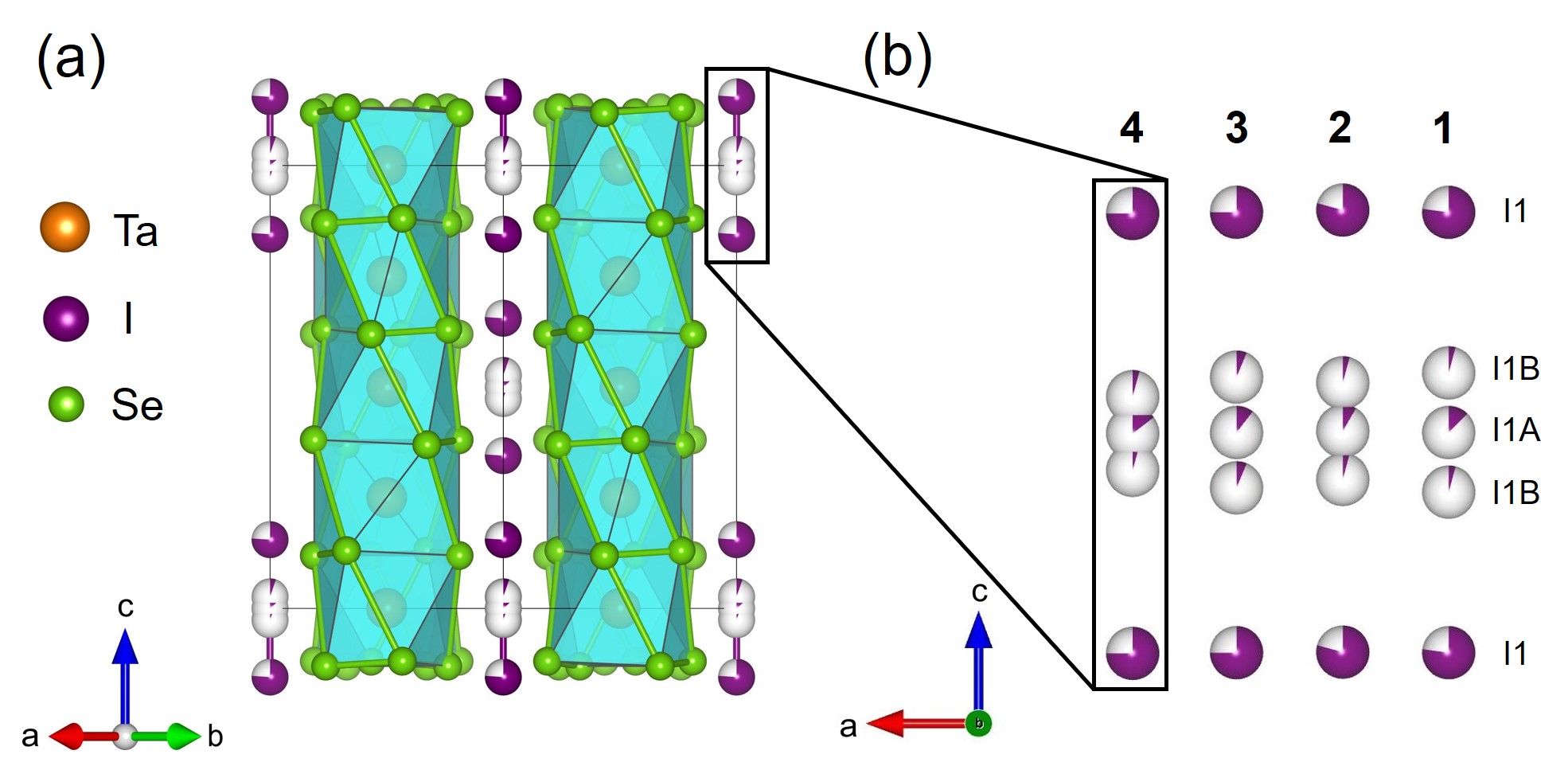} \\
\caption{\label{fig:iodine_deficiency}
Single-crystal XRD refined cells for demonstrating an iodine deficiency. (a) Crystal refinement for sample 4 from Table \ref{table:full_refinement_table} for data collected at $T$ = 100 K. Composition refined to (TaSe$_4$)$_2$I$_{0.873(53)}$. (b) Four separate refinements for three different crystals, demonstrating how the site occupancies have slight variations, but the overall trend is consistent. Each layer of atoms stacked along the $a$ direction belongs to a different crystal refinement (numbered for each sample), with refinemnt parameters given in Table \ref{table:full_refinement_table}. For sample number 4, the I1, I1A, and I1B sites give occupancies of 0.756(6), 0.14(6), and 0.05(3), respectively.
} 
\end{figure}



Our single-crystal XRD confirms the presence of a bulk iodine deficiency in our samples. Refinement parameters from four different full X-ray data sets are given in Table \ref{table:full_refinement_table} (with additional detailed parameters for sample 3 found in the Supplemental Material \cite{supplement}, Tables S2 and S3), while Fig. \ref{fig:iodine_deficiency} displays examples of the unit cells. Although the exact occupancies varied from crystal to crystal, the overall deficiency remained relatively consistent with a weighted average of approximately 0.88(2) total iodine fractional occupancy. Figure \ref{fig:iodine_deficiency}(a) shows the unit cell refinement from sample 4 of Table \ref{table:full_refinement_table} with the associated occupancy information displayed as fractional atomic spheres. Figure \ref{fig:iodine_deficiency}(b) shows four iodine chains from each of the four crystals stacked next to one another, demonstrating the slight variation between refinements for position and fractional occupancy. Refinements 1 and 2 are separate crystals at room temperature, while refinements 3 and 4 are the same crystal examined at room temperature and $T$ = 100 K, respectively. For sample number 4, the regular iodine I1 position gives the highest occupancy of 0.756(6). The I1A site gives an occupancy of 0.14(6), and the I1B site gives an occupancy of 0.05(3). Details about the crystal refinement process are given in the Supplemental Material \cite{supplement}.

This iodine deficiency is likely related to the electrical resistance of (TaSe$_4$)$_2$I, which is well-documented in the vicinity of the CDW temperature \cite{maki_charge_1983, wang_charge_1983, forro_hall_1987, gooth_axionic_2019}, and later found to contain a semiconductor-semiconductor phase transition \cite{tournier-colletta_electronic_2013,kim_kramers-weyl_2021}. Below the transition temperature, the observed temperature dependence is not unexpected due to the opening of band gaps. Above the transition temperature, however, the \tsi\ resistance is in stark contrast to the DFT-calculated band structure of the ideal, high-temperature structure, which places the Fermi level above the Weyl nodes, seeming to imply metallic conductivity. Previous studies stress the importance of the electron-phonon interaction in \tsi\ as a basis for this resistivity behavior \cite{crepaldi_optically_2022, perfetti_spectroscopic_2001}. Vacancies from the iodine deficiency have also been proposed to underlie this discrepancy, along with transverse displacements of Ta ions.\cite{tournier-colletta_electronic_2013} Both could lead to the observed transport behavior. The iodine vacancies likely contribute to the high-temperature disorder that we explore in Section \ref{diffuse}.


With an understanding of the intensity distribution of a scattering pattern, we can determine if our material contains a chiral domain boundary. Friedel's law dictates the existence of indistinguishable points within reciprocal space known as Friedel pairs, points with inverted phase but identical amplitudes. If a crystal has centrosymmetric symmetry the Friedel pairs will have exactly the same amplitudes. When symmetry is broken, the Bragg reflection positions are not altered. Instead, the intensities of Friedel points differ when dynamical absorption effects are considered in crystals containing species of differing atomic numbers. This distinction is utilized in the calculation of the Flack parameter \cite{flack_enantiomorph-polarity_1983,flack_use_2008,parsons_use_2013,watkin_howard_2020,valentin-perez_chirality_2022}, a refinement factor used for determining chirality and given in the form
\begin{equation}
    I(hkl) = (1-x)|F(hkl)|^2+x|F(\bar{h}\bar{k}\bar{l})|^2
\end{equation}
where $F(hkl)$ is the calculated structure factor, $F(\bar{h}\bar{k}\bar{l})$ is the structure factor for the inverted phase, $I(hkl)$ is the observed structure factor, and $x$ is the Flack parameter. The Flack parameter quantifies the correctness of a refinement's absolute structure, reported as a fraction between 0 and 1. A value of 1 indicates a structure which is fully inverted from its correct configuration ($I(hkl) = |F(\bar{h}\bar{k}\bar{l})|^2$), while a value of 0 indicates the proper configuration with no inversion ($I(hkl) = |F(hkl)|^2$). Intermediate values represent different fractions of inversion twinning. For example, a Flack parameter which converges to around 0.5 would suggest the crystal in question has a mixture of 50\% left handed and 50\% right handed configurations.


\begin{figure}[hbt!]
\centering\includegraphics[width=\columnwidth]{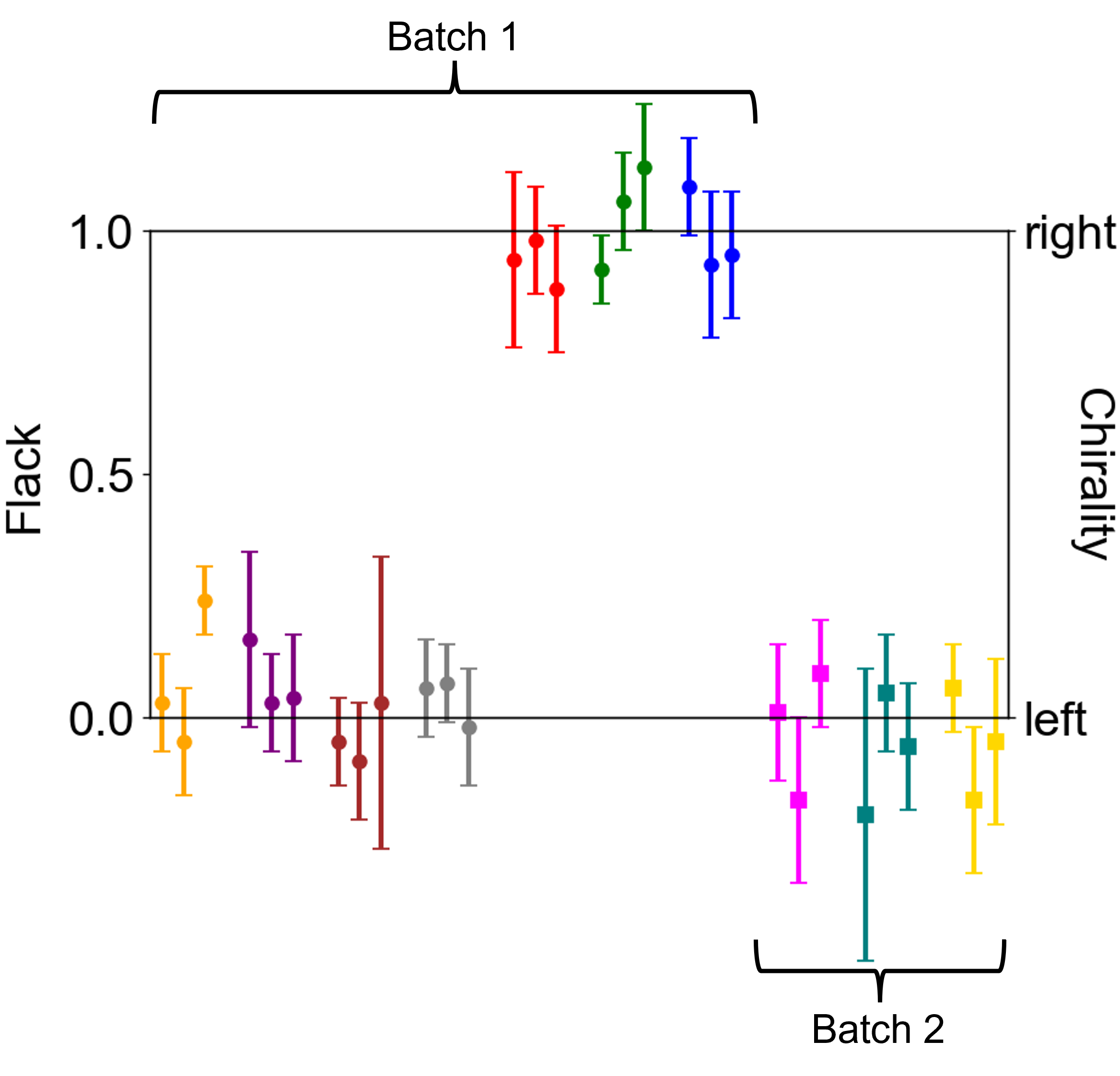} \\
\caption{
\label{fig:Ta2Se8I_handedness}
The Flack parameter and corresponding handedness returned by single-crystal XRD measurements done on 10 separate \tsi\ crystals between two different batches. The colors each designate one of the 10 crystals. Each crystal had three pieces extracted from the middle and ends, giving a total of 30 measurements (three data points per color). The trends shown in this figure suggest that chirality remains consistent within each \tsi\ crystal, but may vary among crystals.
} 
\end{figure}

A previous study from Yi, et al.\cite{yi_surface_2021} reported \tsi\ chirality trends through the use of the Flack parameter in their Supplemental Material. Only one Flack parameter was reported with a value of $\sim$0.3(1), which may suggest inversion twinning. This value was an average value for an undisclosed number of crystals tested by the group. To build upon this work and provide a clearer picture of the \tsi\ chirality trends, we present additional, more detailed studies of the size and distribution of chiral domains through structure refinements done on several locations across multiple large (millimeter-scale) crystals. A total of 10 crystals were tested from two synthesis batches. The crystals were too large to be examined in their as-grown state so the crystals were cut into several pieces. When cut, the crystals delaminate into very thin needles. Three needles from each large crystal were examined for intensity data – one from each end of the crystal and one from the middle section of the crystal for a total of 30 specimens. The resulting integrated intensities were refined against the atomic positions of a reference \tsi\ crystal (sample 3 from Table \ref{table:full_refinement_table}), giving us 30 different Flack parameters. A full scattering data set was previously collected on the reference crystal, showing excellent refinement statistics (see Tables S2 and S3 in the Supplemental Material \cite{supplement}) along with a Flack parameter which converges to near zero, indicating single chirality. Because the reference crystal has a known handedness, refining against this crystal will determine each crystal fragment's handedness with respect to the handedness of the reference crystal. A similar method was used by di Gregorio, et al.\cite{di_gregorio_emergence_2020} In this circumstance, a Flack parameter of 0 indicates a crystal fragment with the same handedness (left) as the reference crystal, and a value of 1 indicates a crystal with opposite handedness (right). The resulting data can be viewed in Fig. \ref{fig:Ta2Se8I_handedness}; additional refinement parameters can be found in Table S4 of the Supplemental Material \cite{supplement}. We found that handedness remained consistent within each crystal, save a few higher uncertainties in measurement; that is, all three pieces from a single crystal shared the same, uniform handedness. However, handedness was observed to vary among separate crystals. 





\begin{figure}
\centering\includegraphics[width=\columnwidth]{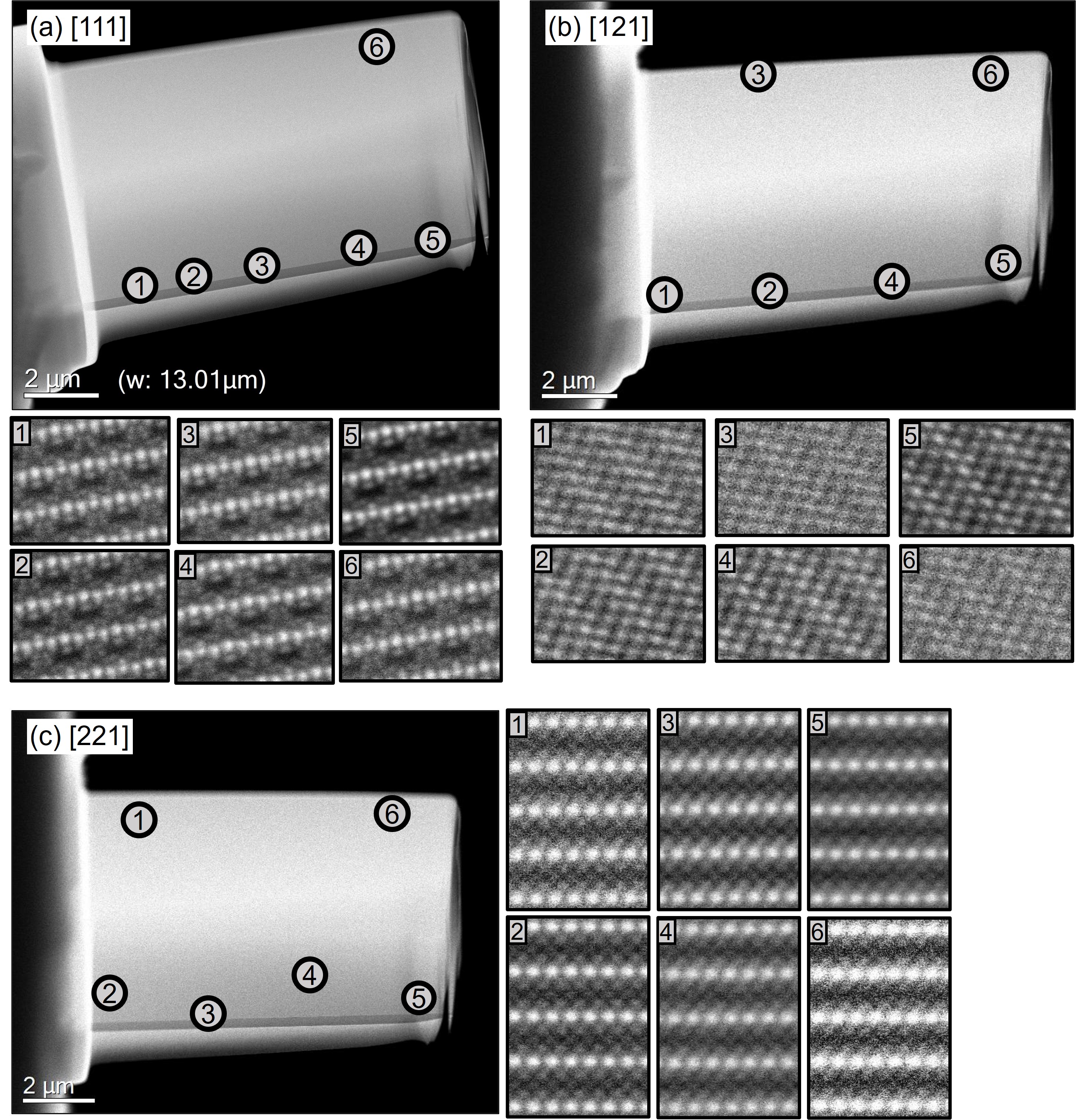} \\
\caption{\label{fig:TSI_Themis}
ADF-STEM imaging shows no change in chirality of a \tsi\ crystal over 10 $\mu$m among several different locations. (a - c) Show six ADF-STEM images each from the [111], [121], and [221] orientations, respectively. The location corresponding to each ADF-STEM image is marked by a number on the large FIB image. The Ta atoms appear brighter than the I atoms, while the I atoms appear brighter than the Se atoms.
} 
\end{figure}

\subsection{Electron microscopy}

Single-crystal XRD enantiopurity is supported by ADF-STEM atomic resolution imaging of a \tsi\ crystal, shown in Fig. \ref{fig:TSI_Themis}. Because STEM images are two-dimensional projections, determining the chirality of a sample via STEM requires imaging from multiple zone axes \cite{dong_atomic-level_2020}. Therefore, ADF-STEM images were acquired of the same grain of \tsi\ from the four zone axes [111], [121], [112], and [121]. Then, for each orientation the \tsi\ crystal was imaged at six different locations which covered a range of roughly 10 $\mu$m. Orientations were observed to vary by no more than 4\degree{} from one another. We attribute this variation to twisting caused by the sample preparation process. Otherwise, all images from their respective zone axes show resolvable, well-stacked, and non-rotated \tsi\ chains. This indicates single chirality over depth (tens of nm) and breadth within the crystal tested.

\subsection{Optical chiral response}
\begin{figure}
\centering\includegraphics[width=0.7\columnwidth]{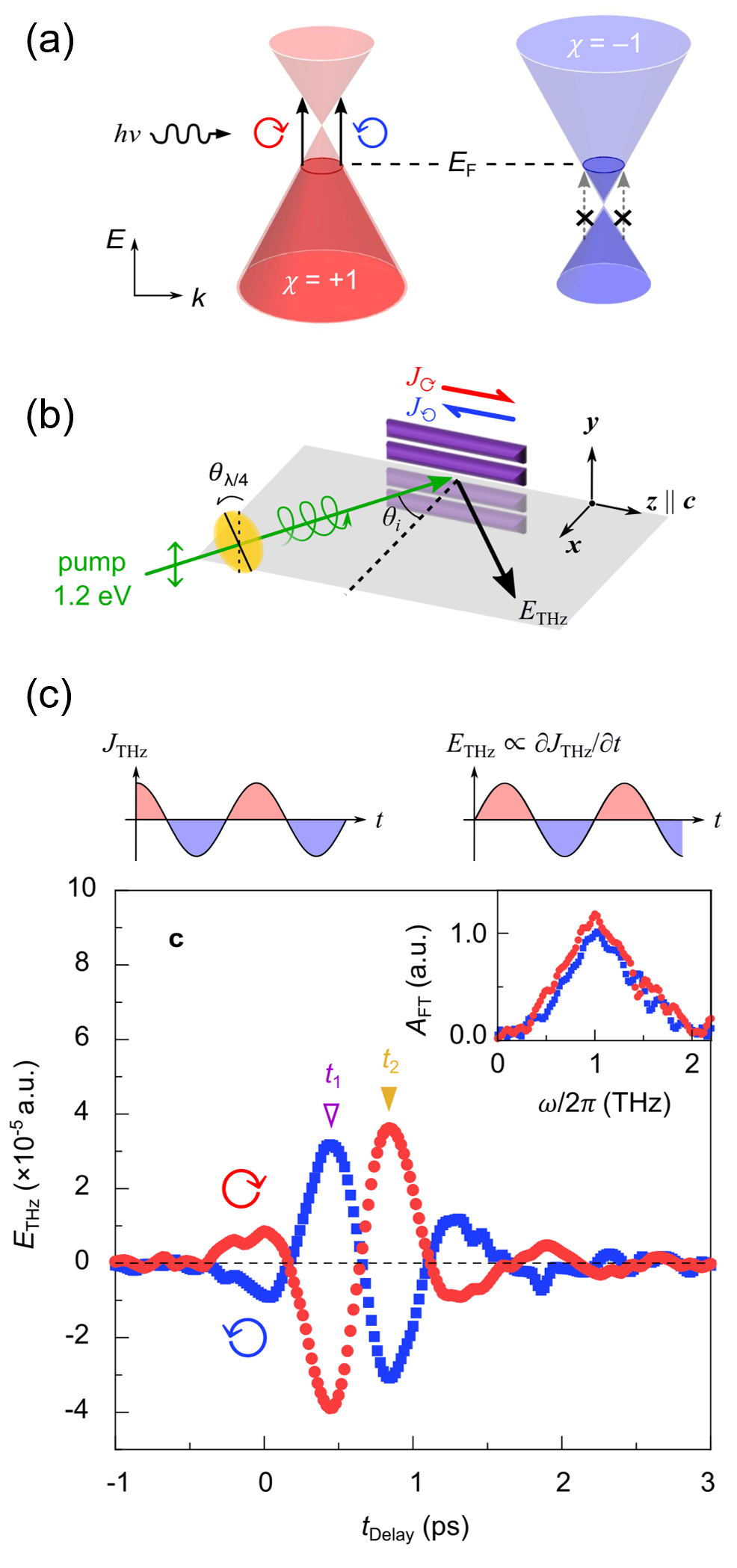} \\
\caption{\label{fig:CPGE}
Helicity-dependent THz emission spectroscopy of (TaSe$_4$)$_2$I. (a) Schematic representation of a pair of Kramers-Weyl (KW) fermions, each positioned above and below the Fermi energy $E$\textsubscript{F}. While the $\chi = +1$ Weyl cone can contribute to the CPGE, the effect is suppressed in $\chi = -1$ due to the Pauli blocking of interband transitions. In Weyl semimetals with mirror symmetry (unlike (TaSe$_4$)$_2$I), Weyl cones are degenerate in energy, and both may contribute to the CPGE which exactly cancels out the effect. (b) THz emission spectroscopy setup and the measurement geometry. The circularly-polarized pump beam (1.2 eV) is
guided onto the sample at $\theta_{i}$ = 45° incident angle. (c) The transient photocurrent $E_\textrm{THz}$ spectra of \tsi\ are plotted by the pump-probe time delay (\emph{t}\textsubscript{Delay}). (Inset) Fourier transform amplitude spectra of the $E_\textrm{THz}(t)$.
} 
\end{figure}

A third confirmation for the consistency of structural chirality in \tsi\ is provided through measuring circular photogalvanic effects (CPGE) using THz emission spectroscopy with a circularly polarized pump \cite{book_photovol}. Since \tsi\ is structurally chiral but non-magnetic (preserves time-reversal symmetry), it naturally contains Kramers-Weyl fermions as illustrated in Fig. \ref{fig:CPGE}(a) \cite{chang_topological_2018}. In such systems, a photogalvanic current can be generated using circularly polarized light. Crucially, opposite helicities of light (right vs. left) will produce currents in opposite directions provided the sample has a constant chirality over the spot size of the light \cite{rees_helicity-dependent_2020}. A contact-free method to measure CPGE is THz emission spectroscopy whereby a short duration IR pulse generates a transient current which emits THz radiation in the far-field (Fig. \ref{fig:CPGE}(b)). The emitted THz can be measured in the time-domain using standard electro-optic sampling (EOS) methods. The change in the phase of the emitted THz upon change in light helicity indicates non-zero CPGE. Figure \ref{fig:CPGE}(c) shows the THz emission signal as a function of EOS delay on \tsi\ for right and left circularly polarized light. The 1.2 eV light was incident upon the sample at an incidence angle of 45° with a spot size of  approximately 1 mm. As can be seen, opposite helicities of light result in opposite phases of the EOS signal indicating a circular photogalvanic effect and confirms the constant structural chirality of \tsi\ over the 1 mm laser spot size.

\subsection{Modeling the CDW phase} \label{scattering}

\begin{figure}
\centering\includegraphics[width=\columnwidth]{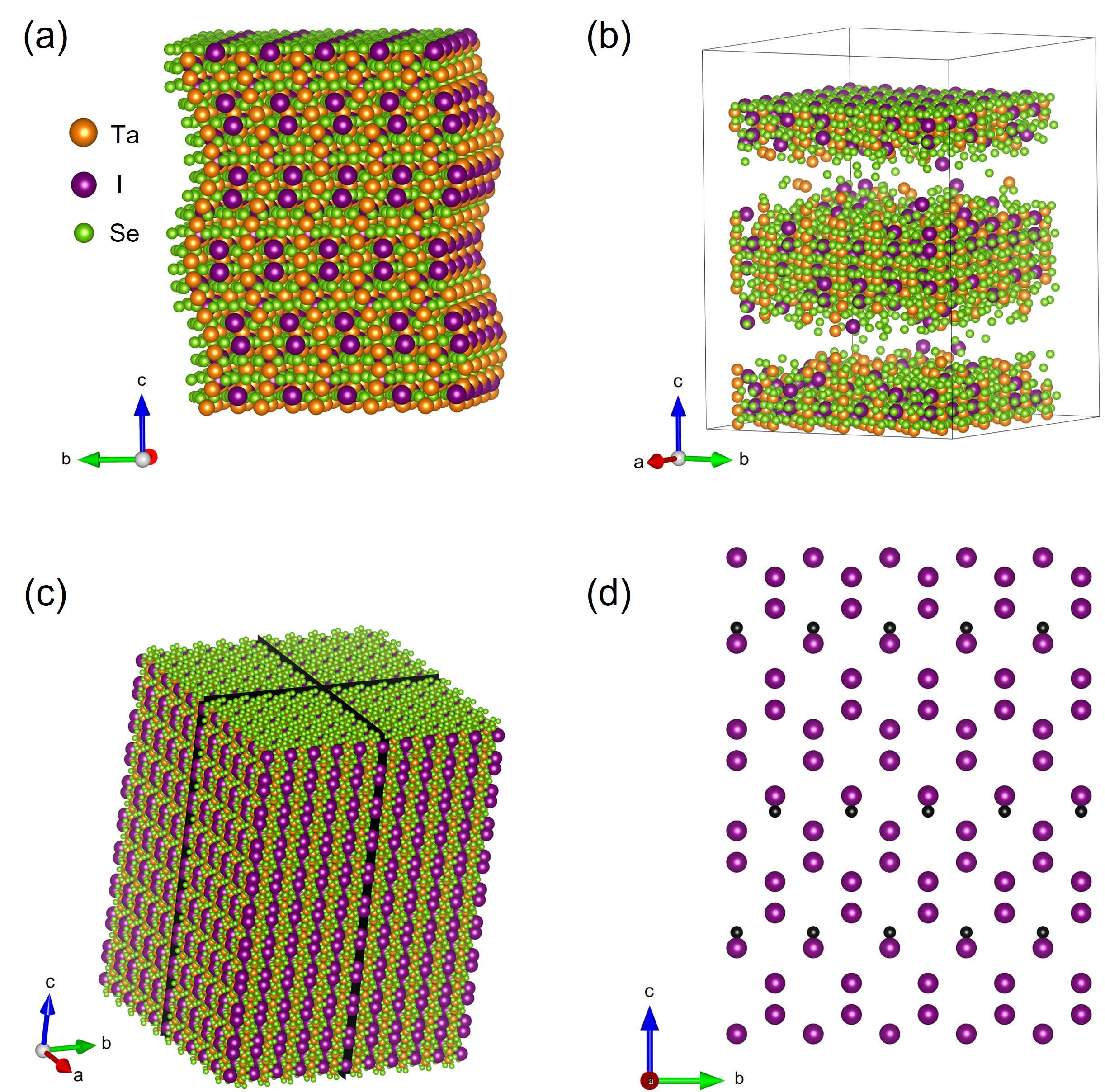} \\
\caption{\label{fig:waves}
(a) Example of a displacement wave applied to all atoms in a \tsi\ lattice with propagation vector along the $c$ direction and the oscillations along the $b$ direction. Displacements are exaggerated for illustration. (b) Example of a density wave applied to all atoms in a \tsi\ lattice with k-vector along the $c$ direction. (c) Domain split used in our diffuse scattering simulations. Domains are all the same size. (d) Model used to simulate the scattering in Fig. \ref{fig:2_plots_L_rods_inset}. Entire $ab$ planes of iodine ions are replaced by voids (black atoms) by random chance, and the nearest iodine ions are shifted towards these voids.
} 
\end{figure}

\begin{figure}
\centering\includegraphics[width=\columnwidth]{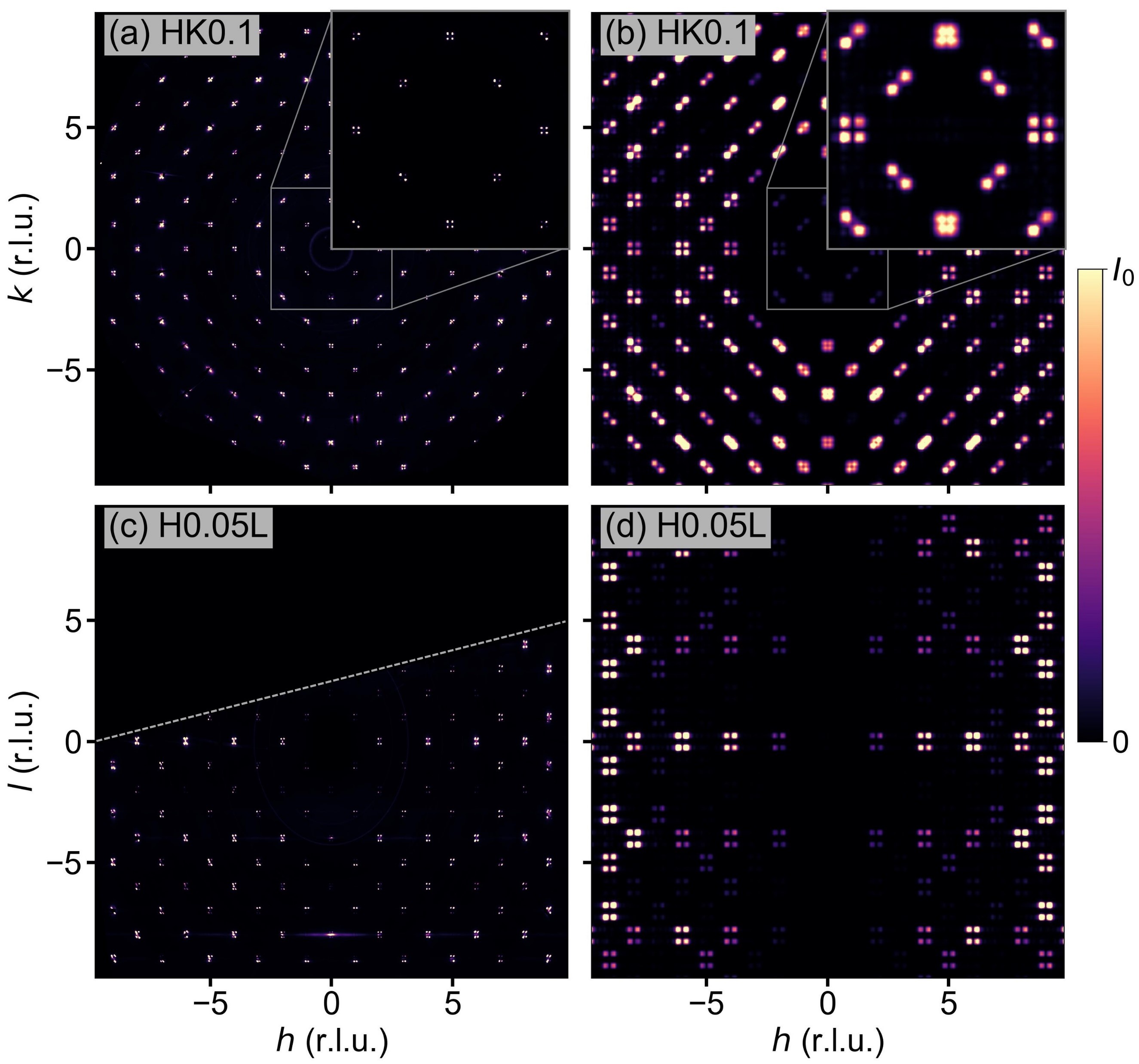} \\
\caption{\label{fig:Paper_CDW_Satellites}
Figures showing views of the CDW satellite peaks for both the experimental and modeled scattering. (a) Experimental scattering measured at $T$ = 30 K in the HK plane at L = 0.1. Four CDW satellite peaks surrounding each Bragg position are visible. The inset displays a magnified view of the marked area. (b) The modeled scattering calculated from a modulated \tsi\ unit cell lattice. Asymmetric intensity patterns are mimicked by confining displacement wave oscillations to the $ab$ plane, orthogonal to the wave vectors. (c) Shows the same experimental scattering of (a) but from the HL plane at K = 0.05. The dotted line marks where the signal cuts off from the geometry of the experimental setup. (d) The simulated scattering at the same plane as (c).
} 
\end{figure}

The single-crystal synchrotron XRD with high dynamic range and better q resolution shows an interesting assortment of diffuse scattering. First-order satellite peaks caused by the CDW phase are readily apparent below the transition temperature. These well-known peaks appear near the Brillouin zone centers in a tetragonal assembly defined by the eight locations $(\pm{\alpha},\pm{\alpha},\pm{\beta}$) rlu, with the main Bragg reflection set as the reference point \cite{fujishita_x-ray_1985, fujishita_incommensurate_1984}. Previous studies have found values ranging from 0.027 to 0.05 for $\alpha$ and 0.012 to 0.151 for $\beta$ \cite{fujishita_neutron_1986, shi_charge-density-wave_2021, smaalen_structure_2001, lee_x-ray_1985, yi_surface_2021, favre-nicolin_structural_2001}. For our crystal, we report numbers of $\alpha \approx 0.05$ and $\beta \approx 0.11$. Lee, et al. speculates the discrepancy in the measured wavevector among studies is correlated with variations in the iodine deficiency of each \tsi\ crystal \cite{lee_x-ray_1985}. This remains an open question, but our measurements of large deficiencies among many crystals supports the viability of a systematic path to confirming if the values vary based on iodine deficiency. Weaker second-order satellites are also found further from the main reflections in our experimental scattering at positions given by $(\pm{2\alpha},\pm{2\alpha},\pm{2\beta}$) \cite{lee_x-ray_1985, nguyen_ultrafast_2022}. The satellites exhibit a large asymmetry in intensities, creating a striking pattern which is observed and well-discussed by Fujishita, et al. and others \cite{fujishita_incommensurate_1984, fujishita_x-ray_1985,monceau_inelastic_1986}.


We seek to build a lattice model modified by static modulations which recreates these satellites through theoretical scattering calculations, requiring knowledge about the number of domains and modulation wavevectors present. A previous comprehensive study done by Van Smaalen, et al. concluded that the \tsi\ CDW phase is a single-$q$, four-domain state, described by a monoclinic supercell with space group $C2$ \cite{smaalen_structure_2001}. Two parts of the modulation wavevector are identified: a stronger transverse acoustic part perpendicular to the chain direction, and a weaker tetramerization of Ta atoms along the chain direction identified as the CDW modulation itself \cite{favre-nicolin_structural_2001, zhang_first-principles_2020}. We attempt a recreation of the transverse modulation by applying four separate but equivalent displacement or density waves to all atoms in a \tsi\ unit cell lattice built in DISCUS. Examples of displacement and density waves are shown in Fig. \ref{fig:waves}(a) and Fig. \ref{fig:waves}(b). To avoid interference between modulation waves leading to mixed higher-order satellites, four equal-sized domains (see Fig. \ref{fig:waves}(c)) are created and each assigned a single modulation wavevector\cite{smaalen_structure_2001}. Due to peak broadening in the simulation, wave vector directions of ($\pm 1$, $\pm 1$, $1$) were chosen to allow more space between satellite peaks so they may remain visually resolvable from one another. We first used density wave modulations, which were applied with a removal probability varying between 0 and 100$\%$. These simulations failed to capture the second-order satellites found at $(\pm{2\alpha},\pm{2\alpha},\pm{2\beta}$), since density waves only create a single pair of peaks around each main reflection per modulation wave. In contrast, displacement waves create infinitely many, equally spaced satellites which fall off in intensity with distance from the main reflections. Therefore, we switched to displacement waves with amplitudes of 0.13 \AA \cite{smaalen_structure_2001}.

Figure \ref{fig:Paper_CDW_Satellites} shows the results of this simulation, demonstrating a model that successfully mimics the asymmetric intensity distribution of the satellites. This was best achieved by confining the displacement wave oscillations to the basal plane and orthogonal to the k-vector direction \cite{lee_x-ray_1985,lorenzo_neutron_1998,smaalen_structure_2001,favre-nicolin_structural_2001}. Otherwise, the satellite peak positions were still correctly simulated, but with incorrect intensity distributions. Although this is not an exact representation of the CDW displacements, we have provided a minimal model that reproduces all of the essential features observed in the data, providing legitimacy for our strategy and additional support for the CDW phase modulations proposed by studies before us.


\begin{figure}
\centering\includegraphics[width=\columnwidth]{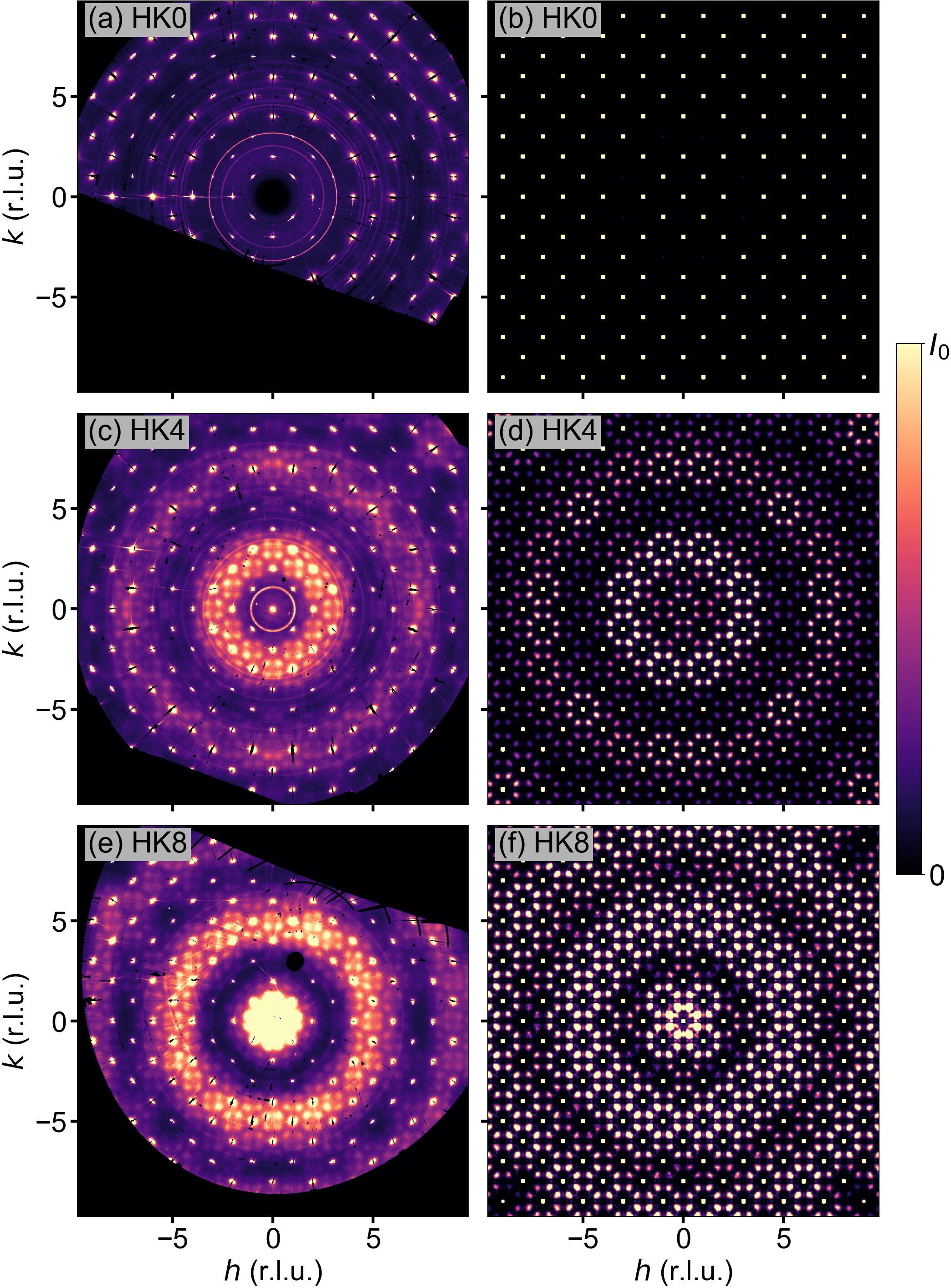} \\
\caption{\label{fig:6_plots_octagon_satellites}
Figures displaying the experimental and simulated octagon diffuse intensity. Three different planes are shown with the experimental scattering at $T$ = 30 K in the left column and the calculated models in the right column. (a, b) Show the HK0 plane, (c, d) show the HK4 plane, and (e, f) show the HK8 plane. Simulated scattering is calculated from a \tsi\ unit cell lattice modulated by displacement wavevectors with directions $(\pm (\sqrt{2}-1), 1, 0)$ and $(\pm 1, \sqrt{2}-1, 0)$. The lack of diffuse intensity at L = 0 means oscillations are along the $c$ direction. Scaling is held constant among all images.
} 
\end{figure}

\subsection{Diffuse scattering} \label{diffuse}

In Section \ref{refinement}, we claim the possibility of local disorder above $T$\textsubscript{CDW} contributing to the elusive \tsi\ transport behavior. Here we report the observation of diffuse satellites far weaker than the CDW satellites which persist both above and below $T$\textsubscript{CDW}, revealed by our high signal-to-noise synchrotron data. Figure \ref{fig:6_plots_octagon_satellites} shows eight diffuse peaks surrounding each Bragg reflection in an octagon-like formation, and are confined within HK planes every four reciprocal lattice units along the L direction. We will refer to these as octagon satellites due to their positioning. Collectively, the octagon satellites form bright, concentric rings which alternate in pattern between each HK plane. The peaks are absent at the HK0 plane, but increase with intensity as L increases.

The methodology we used for simulating the CDW peaks can help elucidate the potential real-space source of the octagon peaks. Considering there are eight peaks around each Bragg reflection, we start again by applying four separate modulation waves to a \tsi\ unit cell lattice, with each modulation assigned to its own equally sized domain. The wavevectors are set in one of four directions from $(\pm (\sqrt{2}-1), 1, 0)$ and $(\pm 1, \sqrt{2}-1, 0)$, which describe the positions of the vertices of an octagon centered on the origin. 

\begin{figure}
\centering\includegraphics[width=\columnwidth]{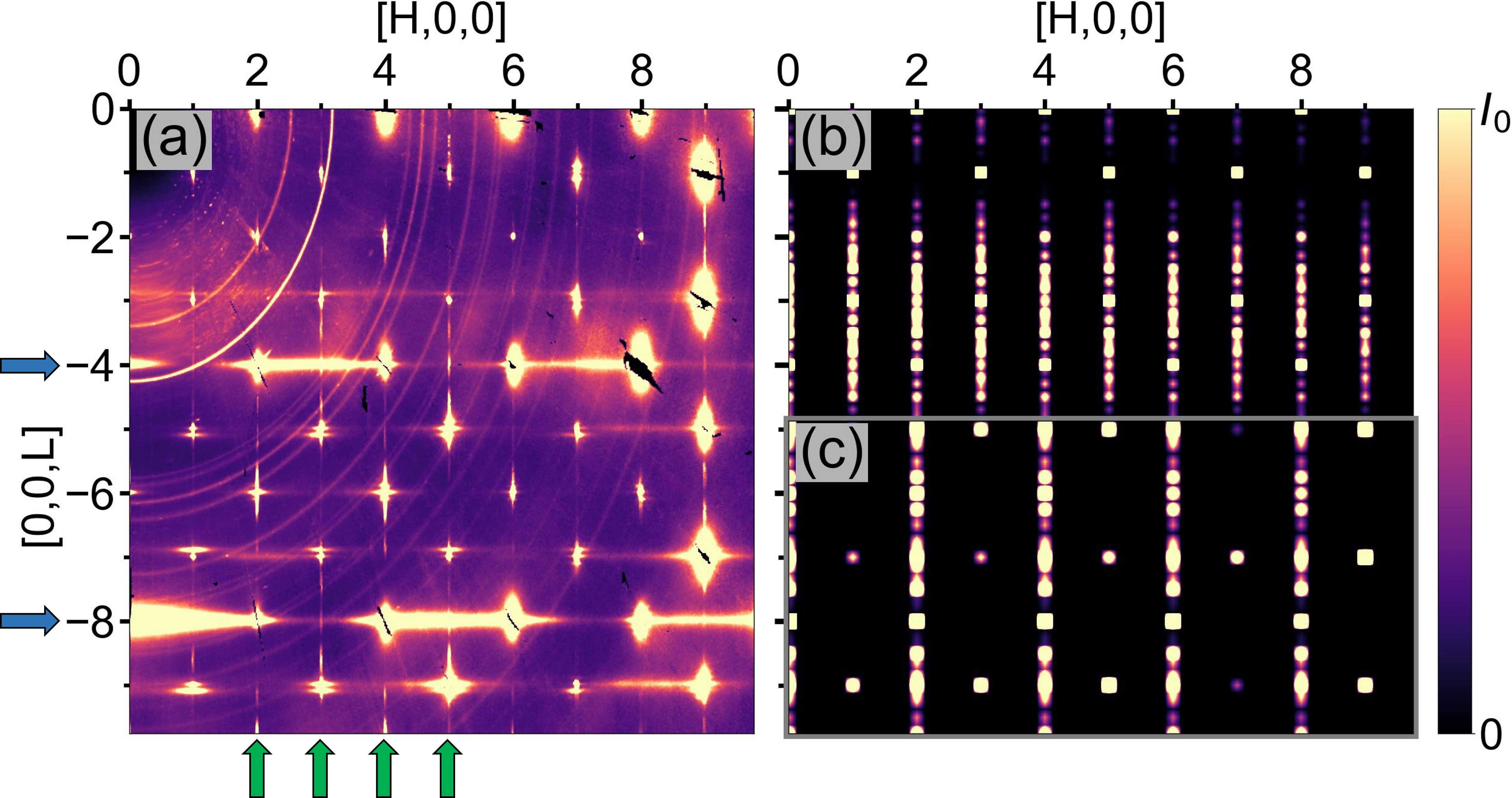} \\
\caption{\label{fig:2_plots_L_rods_inset}
The \tsi\ experimental scattering exhibits diffuse rods running along the L direction (green arrows), as shown in the H0L plane at $T$ = 140 K of figure (a). The intense segmented rods seen along the H direction (blue arrows) mark the locations where the bright rings found in the HK planes (Fig. \ref{fig:6_plots_octagon_satellites}) intersect this H0L plane. (b) Simulated scattering in the H0L plane calculated using a real-space structure similar to Fig. \ref{fig:waves}(d). Only planes of the iodine ions are selected and moved. (c) Same as (b), but with modification of only the Ta atoms. Shifting the iodine ions is necessary to simulate diffuse rods every integer in reciprocal space like that seen in (a).
} 
\end{figure}

Transverse displacement waves with amplitudes of 0.13 \AA~were used in our successful simulations. All atoms were set to oscillate along the $c$ direction, and the wavelength was calculated and chosen to align and overlap neighboring Bragg position satellites. Figure \ref{fig:6_plots_octagon_satellites} shows the results of this simulation. Our model captures the rings of intensity collectively created by the intensity distribution of the octagon satellites. The rings alternate in position along the $c$ direction. Cross sections of the rings are visible as intense, segmented lines oriented along the H or K directions like those seen in Fig. \ref{fig:2_plots_L_rods_inset}. Additionally, the model correctly avoids simulating satellites at the HK0 plane. The intensity scaling among all the plots in Fig. \ref{fig:6_plots_octagon_satellites} remains consistent, showing that the intensity of the octagon satellites increases as we move further from the HK0 plane along the L direction. Remarkably, modulation of the iodine positions is not necessary to obtain the experimental patterns. Any combination of modulations on the three atom types gave the correct satellite positions, but we found that modulating only the Se and Ta atoms together was needed to mimic the intensity distributions. Including the iodine ion gave redundant results, and any other modulation combination failed to simulate the distinct bright rings collectively created by the experimental satellites (Supplemental Material \cite{supplement}, Fig. S5).

We also considered density waves applied to all atoms with removal probabilities ranging between 0 and $100\%$. The results of this simulation are shown in Fig. S6 of the Supplemental Material \cite{supplement}. Calculations from density waves create the correct first-order satellites, but they incorrectly simulate satellite peaks in the HK0 plane which do not increase in intensity along the L direction as seen in the experimental scattering. Transverse displacement waves capture experimental trends more closely because the satellites feature an increase with intensity along the direction of the oscillations, starting from zero intensity and proceeding to infinity as distance from the wave's axis of propagation increases.

\begin{figure}
\centering\includegraphics[width=\columnwidth]{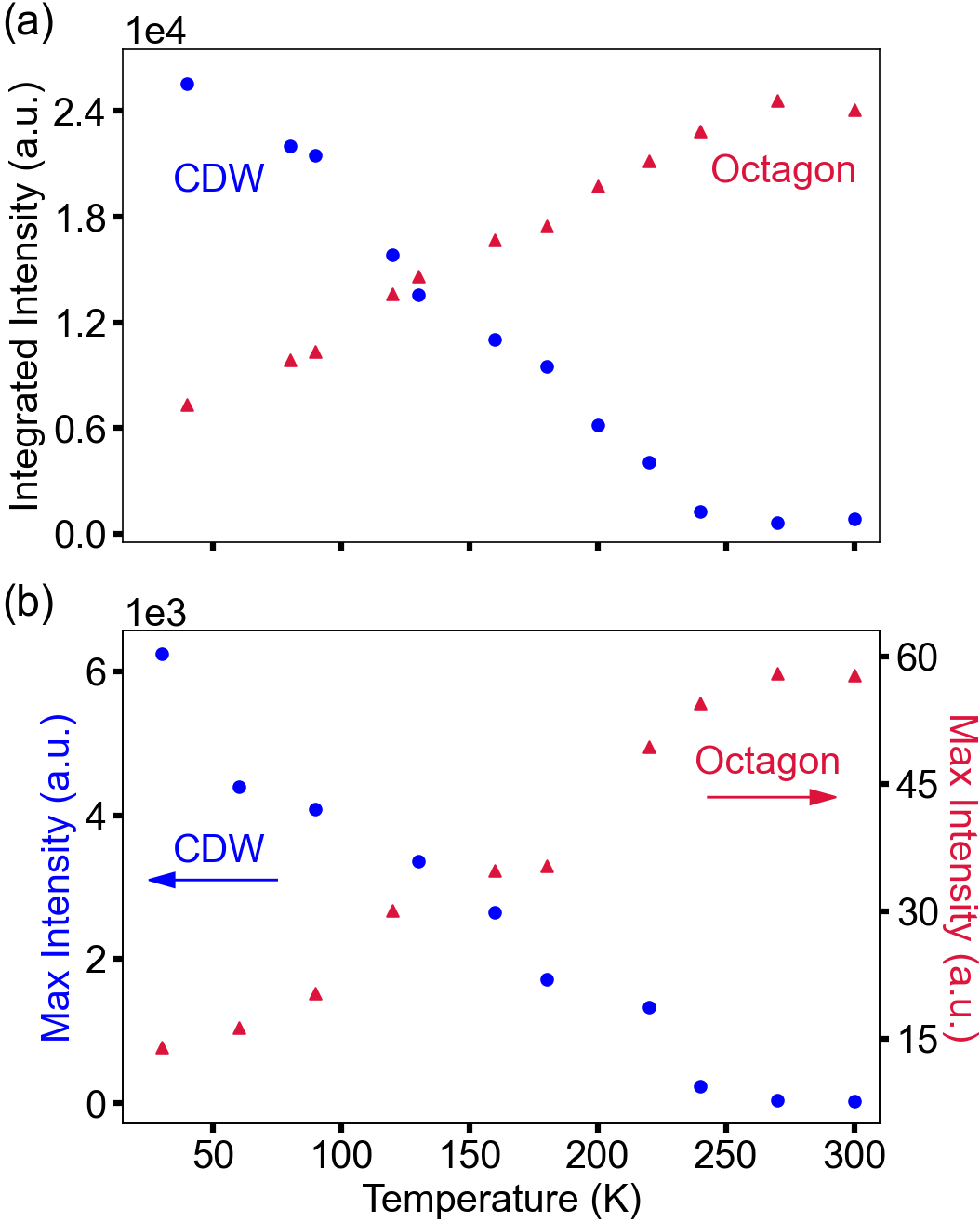} \\
\caption{\label{fig:Tsi_Satellite_Temp}
(a) The integrated intensity variation of a selected CDW and octagon satellite peak with temperature, extracted from the experimental synchrotron X-ray scattering. The CDW and octagon peak coordinates are about $(1.05,2.95,0.11)$ and $(1.3,1.74,4)$ rlu, respectively. Blue circles correspond to CDW peak data, red triangles to octagon peak data. (b) Shows each peak's max intensity variation with respect to temperature. Both figures are background subtracted. The CDW peaks decrease in intensity as temperature increases, tending towards zero as the phase transition is crossed. The octagon peaks, in contrast, steadily increase in intensity as temperature increases.
} 
\end{figure}

A comparison of intensities with respect to temperature between the octagon and CDW peaks can be found in Fig. \ref{fig:Tsi_Satellite_Temp}. Make note that the octagon and CDW integrated intensities are of similar scale, but the maximum intensities differ by a couple orders of magnitude. We notice that the octagon peaks increase in intensity with temperature, and persist above the transition temperature where the CDW peaks do not. Additionally, the octagon peaks tend to subtly shift position within the HK plane and become more diffuse as temperature increases (see Fig. S7 in the Supplemental Material \cite{supplement}). These features alone may suggest that the octagon peaks originate from dynamical (phonon) modulations or electronic coupling to them (polarons).\cite{tournier-colletta_electronic_2013,perfetti_spectroscopic_2001, sinchenko_does_2022} We also notice, however, that the intensities of the peaks level off above the CDW transition temperature. Perhaps the short-range, disordered modulations of the octagon peaks compete with the CDW phase modulations, an explanation for the comparable strength and inverse relationship between the integrated intensities. Further spectroscopic and high-temperature investigation is required.

A separate feature from the displacement-driven octagon modulation is diffuse rods found running along the L direction, repeating every integer in reciprocal lattice units as demonstrated in Fig. \ref{fig:2_plots_L_rods_inset}(a). These diffuse features are present at all temperatures measured, from 30 K to 300 K. We can recreate similar behavior by removing or shifting entire planes of atoms along the $c$ direction (Fig. \ref{fig:waves}(d)). The entire plane must be shifted in order to preserve the ordering in directions perpendicular to the shift. This ensures we only simulate diffuse scattering along a single direction. When only iodine are removed and adjusted, diffuse rods appear in the simulated pattern every integer along H as seen in Fig. \ref{fig:2_plots_L_rods_inset}(b). The simulated diffraction patterns show a stronger intensity of diffuse rods along L than the experimental data, but these simulated rods are still qualitatively similar to the rods marked by the green arrows in the experimental scattering. On the other hand, when tantalum or selenium planes are removed and shifted, diffuse rods are created at every other integer value as seen in Fig. \ref{fig:2_plots_L_rods_inset}(c), so only half of the rods marked with green arrows are accounted for. For this simulation approach, iodine deficiency, and not Ta or Se, leads to the formation of rods at every integer H. The iodine ions are shifted into the highest occupancy positions, the I1A sites found halfway between the regular positions as seen in our refined unit cell (Fig. \ref{fig:iodine_deficiency}(b)). We modeled our simulation in this manner so that it resembles behavior suggested by the iodine occupancies of our crystal refinements. Missing iodine ions cause a relaxation of neighboring atoms towards the gaps, akin to a charge density wave. Unlike the regular, long-range ordering of the CDW phase modulations, the diffuse rods point to a highly disordered shift with short range ordering. Planes to remove and shift are chosen randomly in our simulation, but with a probability that ensures an iodine deficiency of about 10\%, matching the single-crystal XRD refined values.

\section{Conclusions}

We have demonstrated the chirality trends of the one-dimensional Weyl semimetal \tsi\ through single-crystal XRD refinements, ADF-STEM imaging, and helicity-dependent photocurrents. Based on XRD refinement parameters, each separately formed \tsi\ crystal is likely to be enantiopure, while handedness may vary from crystal to crystal within the same synthesis batch conditions.
The formation of chiral domain boundaries in \tsi\ crystals grown by vapor transport is likely rare, and chirality-dependent transport and optical measurements of these materials can likely assume enantiopurity if a preliminary screening is performed via measurement of Flack parameters. 

The single-crystal XRD refinements also revealed a bulk iodine deficiency in our \tsi\ crystals. An average iodine fractional occupancy of 0.88(2) was observed. In general, a bulk iodine deficiency will profoundly impact the band structure of (TaSe$_4$)$_2$I. Defect formation and the subsequent relaxation of the \tsi\ structure can lead to localized electrons, contributing to the semiconductor/insulator resistivity behavior observed in \tsi\ in its high temperature phase, above the formation of a long-range CDW. As a possible footprint of this effect, diffuse scattering streaks can be modeled with concerted motion and disorder of iodine ions along the $c$ direction.

Our real-space pictures of diffuse scattering gives credence to previously hypothesized models for the \tsi\ low temperature phase. The stronger modulation of the CDW phase was simulated using four separate, equal-sized domains, each with their own equivalent transverse displacement wavevector as described by Van Smaalen, et al \cite{smaalen_structure_2001}. Lower-intensity satellites confined to planes orthogonal to the $c$ direction were also observed with high signal-to-noise synchrotron data. We modeled these satellites with additional transverse displacement wavevectors, only requiring modulation of the tantalum and selenium atomic positions. Whether these satellites are the result of dynamical or static modulations is unknown at this point. The positive correlation of intensity with temperature and persistence beyond the CDW phase transition suggests these lower-intensity satellites are the result of dynamical modulations. On the other hand, the intensity trend appears to level off beyond the CDW temperature, and an inverse relationship with the CDW satellite integrated intensities would instead suggest static modulations which compete with the CDW modulations. The results of this study paint a clearer picture of the \tsi\ structure at both low and high temperatures, and open the way for additional real-space studies. 


\begin{acknowledgments}
Sample synthesis, X-ray and THz analysis, and modeling were supported by the Center for Quantum Sensing and
Quantum Materials, an Energy Frontier Research Center
funded by the U.S. Department of Energy (DOE), Office
of Science, Basic Energy Sciences (BES), under Award DE-SC002123.
Synchrotron studies were performed at the Advanced Photon Source, a U.S. DOE Office of Science user facility operated for the DOE Office of Science by Argonne National Laboratory under Contract No. DE-AC02-06CH11357. 
Diffuse scattering modeling was performed on the High Throughput Computing Cluster of the Illinois Campus Cluster, a computing resource that is operated by the Illinois Campus Cluster Program (ICCP) in conjunction with the National Center for Supercomputing Applications (NCSA) and which is supported by funds from the University of Illinois at Urbana-Champaign.
Laboratory X-ray and STEM characterization was carried out in part in the Materials Research Laboratory Central Research Facilities, University of Illinois.
The STEM experiments were supported by the Air Force Office of Scientific Research under Award FA9550-20-1-0302 and by the National Science Foundation under Grant No. 1922758.
We acknowledge additional support from the EPiQS program of the Gordon and Betty Moore Foundation (Grant GBMF11069 for Y.L. and F.M. and Grant GBMF4305 for P.A.).
S.B. acknowledges support through the Early Postdoc.Mobility program from the Swiss National Science Foundation (Grant P2EZP2 191885).
\end{acknowledgments}

\bibliography{main_tsi_diffuse}

\end{document}



\begin{center}
\Large 
\textbf{Disorder and diffuse scattering in single-chirality (TaSe$_4$)$_2$I crystals}\\
\vspace{1em}
Supplementary Material\\
\vspace{1em}
\normalsize
Jacob A. Christensen, Simon Bettler, Kejian Qu, Jeffrey Huang, Soyeun Kim, Yinchuan Lu, Chengxi Zhao, Jin Chen, Matthew J. Krogstad, Toby J. Woods, Fahad Mahmood, Pinshane Y. Huang, Peter Abbamonte, and Daniel P. Shoemaker

\vspace{2em}
\end{center}

\section{Lattice modeling challenges}
\begin{figure}
\centering\includegraphics[width=\columnwidth]{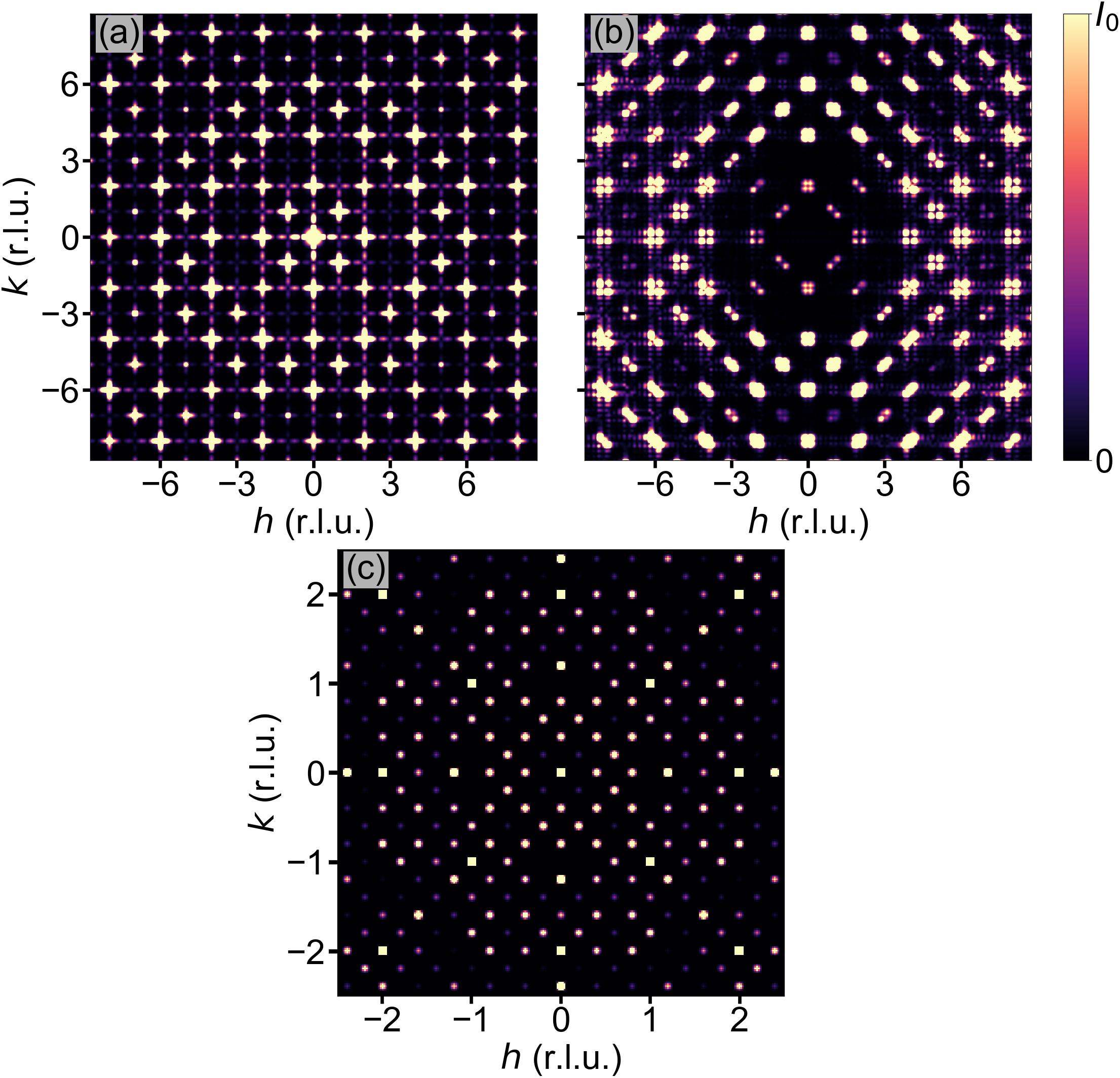} \\
\caption{\label{fig:challenges}
(a) Size effects appear as periodic intensity found along the directions with finite length in a lattice. The periodic structure factor generates intermittent intensity between Bragg positions which scales in strength inversely with the number of cells along a particular direction. An infinite lattice will approach the limit where size effects are no longer observed. When the reciprocal lattice spacing is set to include points only where the periodic structure factor equals zero, size effects can be avoided in a finite lattice simulation. (b) Shows an example of the scattering caused by domain boundaries, interspersed among the simulated CDW satellite peaks. Because the displacement waves creating the satellites are well-ordered, significant subsidiary peaks are seen emanating from each sharp satellite in the directions of the domain boundaries. (c) displays the higher order scattering which becomes more prominent as lattice size is increased. Here, the lattice size was set to 20$\times$20$\times$20 unit cells rather than the 10$\times$10$\times$10 used throughout the study. 
} 
\end{figure}

Computational limitations imposed constraints on the size of our lattices in DISCUS, which raises concerns about edge effects and poses a challenge for generating modulations with long range periodic order. Fortunately, DISCUS software provides the option to enable periodic boundary conditions. Although this removes scattering intensity from the edges of the lattice, finite size effects are still visible as a mathematical consequence of the DISCUS algorithm (see Fig. \ref{fig:challenges}(a)). DISCUS calculates scattering intensity based on the standard kinematical equation, a Fourier transform involving the positions of each species and associated periodic atomic scattering factors. The size effects appear between Bragg reflections from this periodicity, gradually decreasing in intensity as the lattice increases in size. We can entirely omit these periodic intensities along a specific axis by setting the reciprocal lattice spacing equal to $1/N$, where $N$ is the number of unit cells along that axis. Now the scattering will only be calculated where the periodic scattering factor for the perfect lattice is zero, save for the Bragg positions \cite{Neder_Proffen_2008}.

Besides the inherent size effects in DISCUS, we introduce additional unwanted scattering through the domain boundaries we place between modulations. These domain boundaries are needed to avoid additional periodicities created by interference between modulations; however, when left unattended, these boundaries cause periodic scattering intensities to appear perpendicular to the direction the boundaries stretch along (Fig. \ref{fig:challenges}(b)). These scattering effects were lessened in our simulations by applying damping functions to each displacement modulation \cite{Neder_Proffen_2008}. By applying a Gaussian-shaped envelop to the modulating functions, we can damp the amplitude of displacements near domain boundaries, smoothing out an otherwise abrupt transition between domains. The result is a smoothing of scattering features that would otherwise be sharp and visible. The drawback of this method is a reduction in scattering intensities from the modulations due to an overall decrease in the average displacement distance of each atom. When the damping function is too strong relative to the size of the domains, the scattering features may become too weak or diffuse to interpret.

DISCUS provides its users with the option to calculate scattering from the average of smaller sections of a large crystal, known as lots. This ensures that the program will not incorrectly assume that all atoms coherently scatter with one another, as only those correlation lengths that are within the temporal and lateral coherence of the radiation should lead to coherent scattering \cite{Neder_Proffen_2008}. For our simulations, we opted to calculate the complete crystal all at once. This is because our lattice sizes were relatively small, lessening the effects of unwanted long-range coherent scattering. Additionally, our relatively long-range modulations are poorly described by the small volumes used during a lots simulation; the behavior of each modulation cannot be fully realized over the lot, leading to highly diffuse, smeared-out scattering patterns.

DISCUS allows users to subtract a portion of the average scattering away from a calculation, with the assumption that only the diffuse scattering of a disordered structure will remain. This method is also an alternative way to remove finite size effects, assuming all significant size effects are the results of the average structure, and the diffuse intensities are small enough to produce insignificant size effects. Because our modulations are well-ordered over long ranges, this method accomplishes little for us. Sharp satellites are produced by our well-ordered defects, contributing significantly to the average scattering and producing significant subsidiary peaks (Fig. \ref{fig:challenges}(b)). Therefore, we rely on precise grid spacing to omit finite size effects entirely.

\section{Crystal refinement details}

This section addresses the methodology used to refine the \tsi\ crystal structures. For the four fully refined samples with refinement parameters shown in Table \ref{table:full_refinement_table_suppl}, the iodine ion was modeled as disordered over three positions. See these three positions in the unit cell shown in Fig. \ref{fig:sample3_cell}. The majority of 
the time, the iodine atom resides at a 4e position (I1). With only one iodine 
position assigned, there is significant residual electron density at the 2a 
position that was assigned as I1A. Attempts to refine the iodine atoms 
anisotropically with only two assigned iodine positions resulted in 
displacement parameters for I1A that were highly elongated along the $c$ axis 
direction. Reasonable displacement parameters for I1A can be achieved by 
moving that atom slightly off of the 2a special position. However, the 
refinement will not converge with this model. The 2a position is an intersection of a four-fold rotation axis and several perpendicular two-fold rotation axes. During each refinement cycle, the I1A atom would oscillate around to different quadrants of the four-fold axis, occasionally showing large positional shifts along the $c$ axis direction. Convergence of the I1A position was possible by leaving it exactly on the 2a position, which then left a significant residual electron density peak. The final model assigns a 
third iodine position at the 4e position (I1B) to this residual electron density, in close proximity to I1A. This model converges easily with reasonable displacement 
parameters for all atoms. The site occupancy of the three iodine positions 
was allowed to freely refine with similar displacement amplitudes (esd 0.01) 
imposed on disordered sites overlapping by less than the sum of van der Waals 
radii.

\begin{center}

\begin{table}[hbt!]
    \caption{\label{table:full_refinement_table_suppl}
    Same as Table 1 from the main manuscript. Parameters from four full data set single-crystal X-ray refinements. Samples 1 and 2 are separate crystals at room temperature, while sample 3 and 4 are the same crystal measured at room temperature and $T = 100$ K, respectively. Each crystal has an associated handedness revealed by the refinements. Goodness of fit (GooF) measures the quality of our model. A value of 1 indicates a high-quality refinement. The Flack parameter quantifies the confidence in a refinement's absolute structure. A value near zero indicates a proper configuration with no inversion. Each crystal's iodine deficiency is given in terms of the fraction of full occupation (final column).}
        \centering
        \small
        \begin{tabular}{ccccc}
        \hline
            \textbf{Sample} & \textbf{Handedness} & \textbf{GooF} & \textbf{Flack} & \textbf{Occupancy} \\ \hline
            1 & left & 1.149 & 0.017(26) & 0.886(22) \\ 
            2 & left & 1.071 & -0.012(74) & 0.888(50) \\ 
            3 & left & 1.073 & 0.015(24) & 0.882(24) \\ 
            4 & left & 1.235 & 0.021(15) & 0.873(53) \\ \hline
        \end{tabular}
    
\end{table}

\begin{table}[hbt!]
    \caption{\label{table:detailed_parameters}
    Additional refinement parameters for sample 3 from Table \ref{table:full_refinement_table_suppl}.}
        \centering
        \begin{tabular}{c c}
        \hline
            $T$ (K) & 297 \\
            Radiation & MoK$\alpha{}$ \\
            Crystal size (mm) & 0.018 $\times$ 0.063 $\times$ 0.388 \\
            Crystal density (g cm$^{-3}$) & 6.317 \\
            $\mu$ (mm$^{-1}$) &  46.16 \\
            Space group & $I422$ \\
            $a$ (\AA) & 9.5373(9) \\
            $b$ (\AA) & 9.5373(9) \\
            $c$ (\AA) & 12.770(2) \\
            $V$ (\AA$^3$) & 1161.6(2) \\
            $Z$ (K) & 4 \\
            $R$\textsubscript{int} & 0.0477 \\
            $R$\textsubscript{1} & 0.0193 \\
            $wR$\textsubscript{2} & 0.0311 \\
            Total reflections & 8810 \\
            Unique reflections & 1126 \\
            Parameters & 37 \\
            Extinction coefficient (\AA$^{-3}$) & 0.00057(3) \\
            $\Delta \rho$\textsubscript{max}, $\Delta \rho$\textsubscript{min} (e \AA$^{-3}$) & 0.96, -1.59 \\
           \hline
        \end{tabular}

\end{table}

\begin{table}[hbt!]
    \caption{\label{table:displacement_param}
    Refined atomic positions and displacement parameters for sample 3 from Table \ref{table:full_refinement_table_suppl}.}
    \begin{adjustbox}{width=\columnwidth,center}
        \centering
        \begin{tabular}{c c c c c c c c c c c c}
        \hline
            atom & site & x & y & z & $U$\textsubscript{11} (\AA$^2$) & $U$\textsubscript{22} (\AA$^2$) & $U$\textsubscript{33} (\AA$^2$) & $U$\textsubscript{23} (\AA$^2$) & $U$\textsubscript{13} (\AA$^2$) & $U$\textsubscript{12} (\AA$^2$) & $U$\textsubscript{eq} (\AA$^2$)\\ \hline
            Ta1 & 4c & 0 & 0.5 & 0 &  0.0137(2) & 0.0151(2) & 0.0241(2) & 0 & 0 & 0 &  0.0176(1) \\
            Ta2 & 4d & 0 & 0.5 & 0.25 & 0.0141(2) & 0.0141(2) & 0.0237(2) & 0 & 0 & 0.0005(2) & 0.0173(1) \\
            I1$^a$ & 4e & 0 & 0 & 0.15513(8)  & 0.0256(4) & 0.0256(4) & 0.0372(5) & 0 & 0 & 0  & 0.0295(3) \\
            I1A$^b$ & 2a & 0 & 0 & 0 & 0.019(4) & 0.019(4) & 0.04(1) & 0 & 0 & 0 & 0.027(5) \\
            I1B$^c$ & 4e & 0 & 0 & 0.039(4) & 0.030(6) & 0.030(6) & 0.05(2) & 0 & 0 & 0 & 0.037(6) \\
            Se1 & 16k & 0.12156(5) & 0.31228(5) & 0.11921(4) & 0.0183(3) & 0.0168(3) & 0.0256(2) & 0.0009(2) & 0.0023(2) & 0.0021(2) & 0.0202(1) \\
            Se2 & 16k & 0.21620(5) & 0.54445(6) & 0.13052(4) & 0.0160(2) & 0.0200(3) & 0.0252(2) & 0.0018(2) & -0.0010(2) & -0.0021(2) &  0.0204(1) \\
            
           \hline
        \end{tabular}
    \end{adjustbox}

\end{table}

\begin{table}[hbt!]
    \caption{\label{table:parameter_table}
    Table of parameters for each \tsi\ crystal fragment refinement. A partial data set was collected on each sample, which was then refined against the atomic coordinates of a reference crystal (sample 3 of Fig. \ref{table:full_refinement_table_suppl}). The \emph{Crystal reference} column refers to each of the ten large crystals tested. \emph{Location} refers to where each \tsi\ piece was extracted from the larger crystal. Goodness of fit measures the quality of our model. A value of 1 indicates a high-quality refinement. The Flack parameter quantifies the confidence in a refinement’s absolute structure. A value near zero indicates a proper configuration with no inversion. Each crystal’s iodine deficiency is given in terms of the fraction of full occupation (final column). The data of this table suggest that consistent chirality is found within each of the ten large crystals. All pieces from a particular crystal return the same chirality for that crystal. Handedness may vary among different crystals.}
    \begin{adjustbox}{width=\columnwidth,center}
        \centering
        \begin{tabular}{|c|c|c|c|c|c|c|}
        \hline
            \textbf{Sample number} & \textbf{Crystal reference} & \textbf{Location} & \textbf{Handedness} & \textbf{Goodness of fit} & \textbf{Flack parameter} & \textbf{Occupancy} \\ \hline
            \multicolumn{7}{|c|}{\textbf{Batch 1}} \\ \hline
            6205d & xtal 1 & end & right & 0.972 & 0.06(18) & 0.879(48) \\ \hline
            6205d1 & xtal 1 & end & right & 0.974 & 0.02(11) & 0.859(36) \\ \hline
            6205d2 & xtal 1 & middle & right & 0.996 & 0.2(2) & 0.91(28) \\ \hline
            6205d3 & xtal 2 & end & right & 0.968 & 0.08(7) & 0.882(34) \\ \hline
            6205d4 & xtal 2 & end & right & 1.012 & -0.06(10) & 0.879(26) \\ \hline
            6205d5 & xtal 2 & middle & right & 1.059 & -0.07(17) & 0.922(92) \\ \hline
            6205d6 & xtal 3 & end & right & 1.038 & -0.09(10) & 0.882(45) \\ \hline
            6205d7 & xtal 3 & end & right & 1.102 & 0.07(15) & 0.928(37) \\ \hline
            6205d8 & xtal 3 & middle & right & 1.01 & 0.05(13) & 0.885(40) \\ \hline
            6205d9 & xtal 4 & end & left & 1.153 & 0.03(10) & 0.900(38) \\ \hline
            6205d10 & xtal 4 & end & left & 1.335 & -0.05(11) & 0.892(55) \\ \hline
            6205d11 & xtal 4 & middle & left & 1.027 & 0.21(9) & 0.893(35) \\ \hline
            6205d12 & xtal 5 & end & left & 1.039 & 0.16(18) & 0.962(46) \\ \hline
            6205d13 & xtal 5 & end & left & 1.099 & 0.03(10) & 0.88(23) \\ \hline
            6205d14 & xtal 5 & middle & left & 1.461 & 0.04(13) & 0.859(38) \\ \hline
            6205d15 & xtal 6 & end & left & 1.032 & -0.05(9) & 0.877(23) \\ \hline
            6205d16 & xtal 6 & end & left & 1.04 & -0.09(12) & 0.900(23) \\ \hline
            6205d17 & xtal 6 & middle & left & 0.978 & 0.03(30) & 0.911(30) \\ \hline
            6205d18 & xtal 7 & end & left & 1.124 & 0.06(10) & 0.873(41) \\ \hline
            6205d19 & xtal 7 & end & left & 1.049 & 0.07(8) & 0.882(19) \\ \hline
            6205d20 & xtal 7 & middle & left & 1.11 & -0.02(12) & 0.890(38) \\ \hline
            \multicolumn{7}{|c|}{\textbf{Batch 2}} \\ \hline
            6205d21 & xtal 8 & end & left & 1.346 & 0.01(14) & 0.924(39) \\ \hline
            6205d22 & xtal 8 & end & left & 1.27 & -0.17(17) & 0.885(70) \\ \hline
            6205d23 & xtal 8 & middle & left & 1.016 & 0.09(11) & 0.95(50) \\ \hline
            6205d24 & xtal 9 & end & left & 1.527 & -0.2(3) & 0.894(32) \\ \hline
            6205d25 & xtal 9 & end & left & 0.971 & 0.05(12) & 0.869(34) \\ \hline
            6205d26 & xtal 9 & middle & left & 1.181 & -0.06(13) & 0.95(20) \\ \hline
            6205d27 & xtal 10 & end & left & 1.124 & 0.06(9) & 0.905(23) \\ \hline
            6205d28 & xtal 10 & end & left & 1.307 & -0.17(15) & 0.919(56) \\ \hline
            6205d29 & xtal 10 & middle & left & 1.502 & -0.05(17) & 0.881(69) \\ \hline
        \end{tabular}
    \end{adjustbox}

\end{table}

\begin{figure}[hbt!]
\centering\includegraphics[width=0.6\columnwidth]{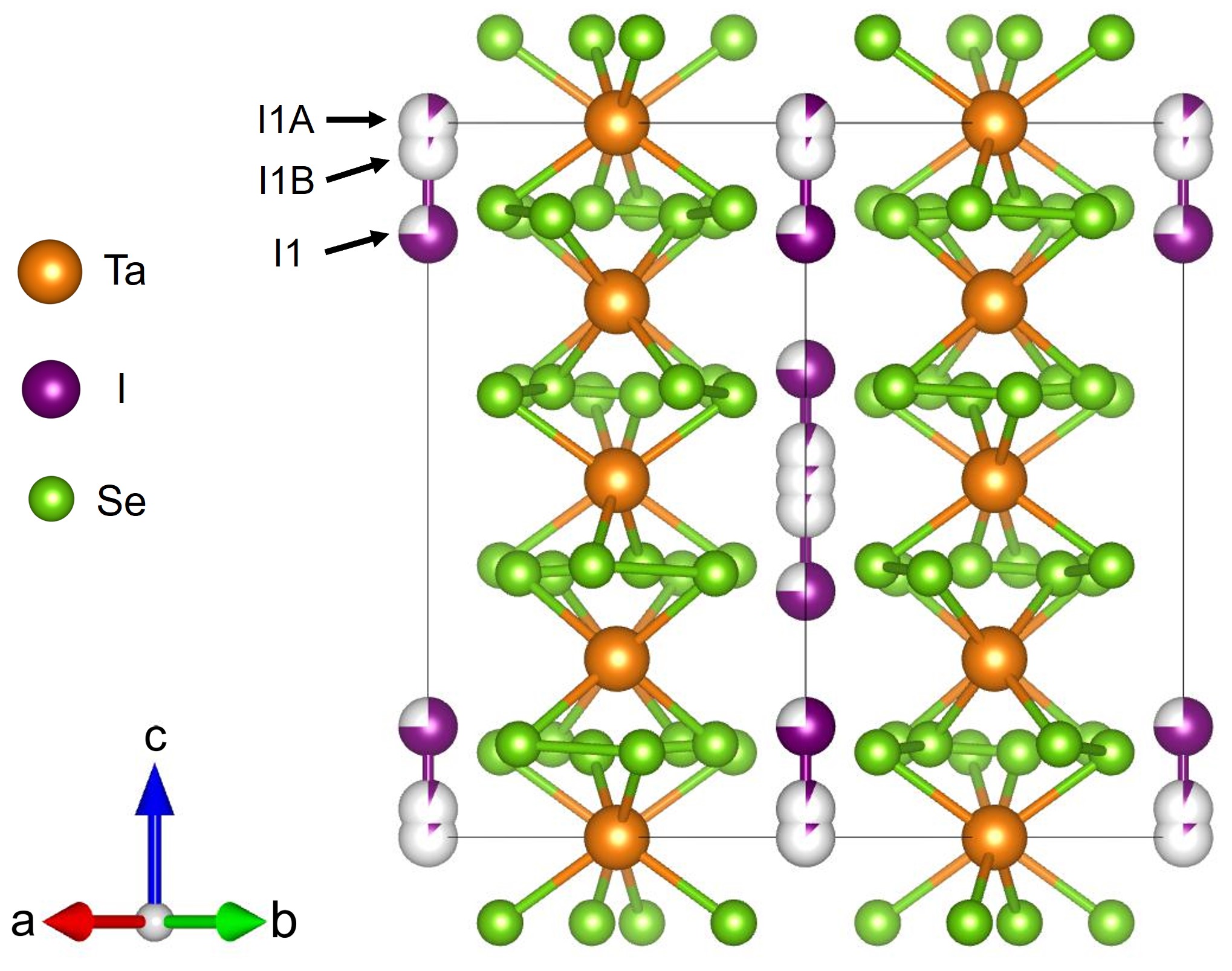} \\
\caption{\label{fig:sample3_cell}
The unit cell refinement for sample 2 of Table \ref{table:full_refinement_table_suppl}. The iodine ions were modeled over the three locations as marked. Occupancy information is shown as fractional atomic spheres. 
} 
\end{figure}

\begin{figure}[hbt!]
\centering\includegraphics[width=0.82\columnwidth]{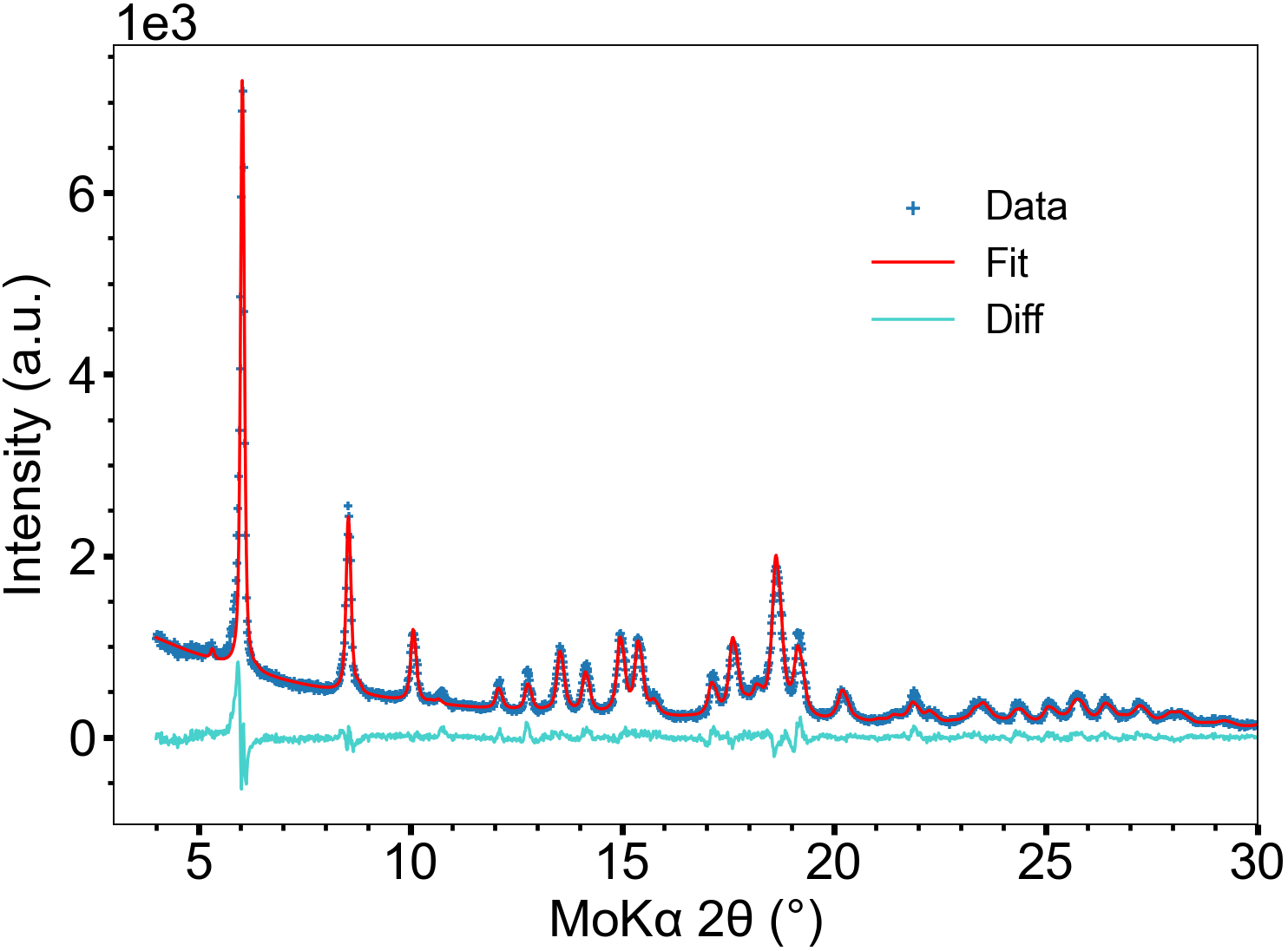} \\
\caption{\label{fig:Ta2Se8I_Powder_Fit}
The powder X-ray diffraction refinement for one of our \tsi\ crystals. \tsi\ powder was loaded into a capillary for data collection on a Bruker D8 Advance with MoK$\alpha$ radiation. Rietveld refinements match the high temperature $I422$ phase, confirming phase purity.
} 
\end{figure}

\begin{figure}
\centering\includegraphics[width=0.82\columnwidth]{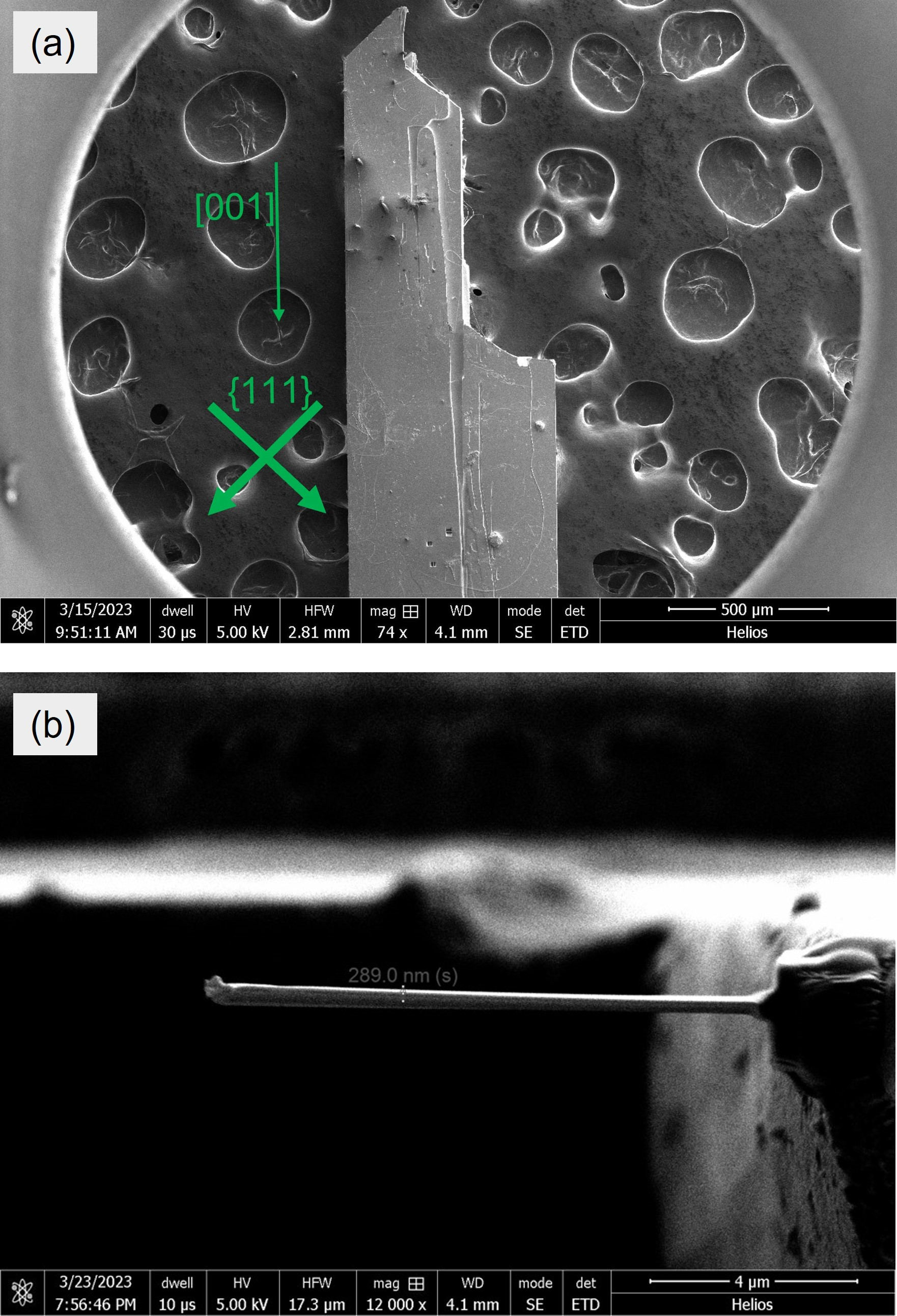} \\
\caption{\label{fig:additional_TEM1}
Micrographs of a (TaSe$_4$)$_2$I sample mounted for focused-ion beam liftout (a) and after milling into a cantilever for analysis by transmission electron microscopy, viewed in cross section in (b).
} 
\end{figure}

\begin{figure}
\centering\includegraphics[width=0.82\columnwidth]{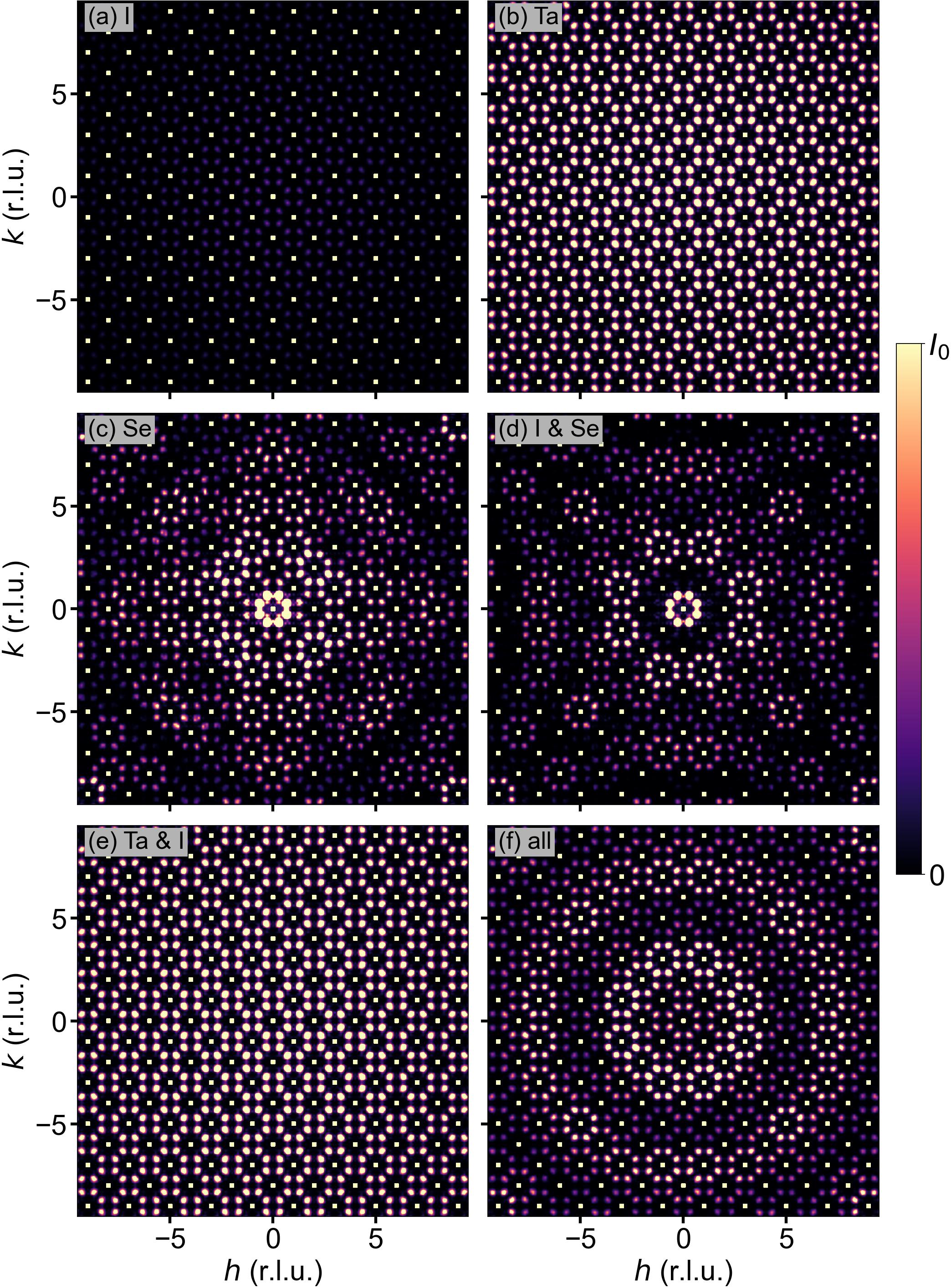} \\
\caption{\label{fig:shift_atom_types}
Figures showing the simulated octagon satellites created from displacement waves only shifting the atom types as follows: (a) iodine, (b) tantalum, (c) selenium, (d) iodine and selenium, (e) tantalum and iodine, (f) all atoms. All frames show scattering on the HK4 plane. Frame (f) exhibits the same intensity pattern as the results shown in the main paper (where displacement waves only modulate tantalum and selenium). This suggests that the modulation of the iodine positions is unnecessary to produce satellites observed in the experimental scattering. All other frames produce incorrect intensity patterns.
} 
\end{figure}

\begin{figure}
\centering\includegraphics[width=0.77\columnwidth]{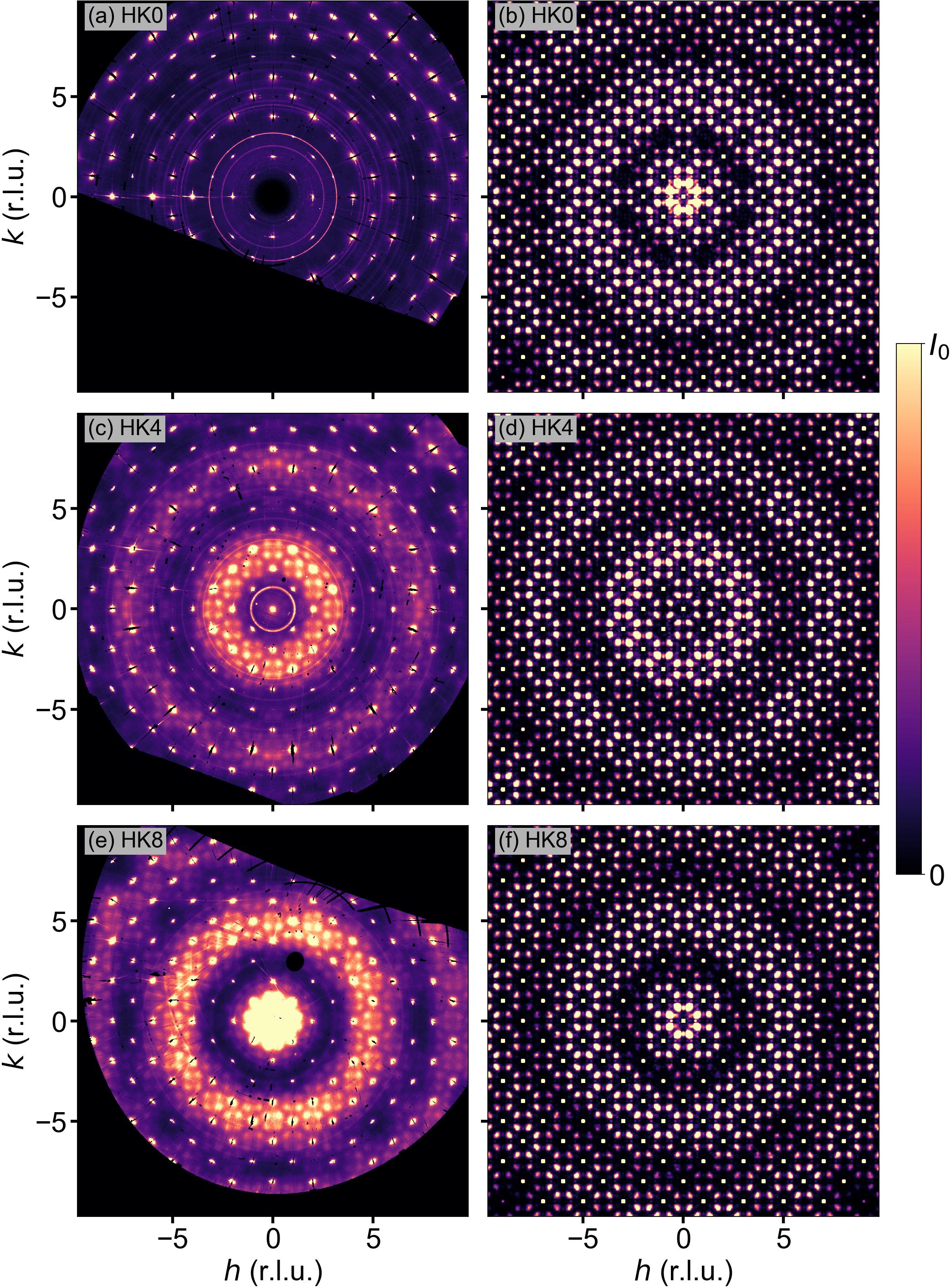} \\
\caption{\label{fig:6_plots_octagon_satellites_density}
Figures displaying the experimental and simulated octagon satellite peaks using density waves. Density waves are applied to all atoms, with directions $(\pm (\sqrt{2}-1), 1, 0)$ and $(\pm 1, \sqrt{2}-1, 0)$, which are the coordinates for the vertices of an octagon. Wave amplitudes (probability of removing atoms) varies between 0 and 100$\%$. Three different planes are shown with the experimental scattering at $T = 30$ K in the left column and the calculated model in the right column. (a, b) Show the HK0 plane, (c, d) show the HK4 plane, and (e, f) show the HK8 plane. Simulated scattering is calculated from a modulated $10\times10\times10$ \tsi\ unit cell lattice. Scaling is held constant among all images. For this simulation, density waves properly simulate the satellite positions and the rings of intensity, but incorrectly simulate diffuse scattering at the HK0 plane where none is observed in the experimental.} 
\end{figure}

\begin{figure}
\centering\includegraphics[width=0.82\columnwidth]{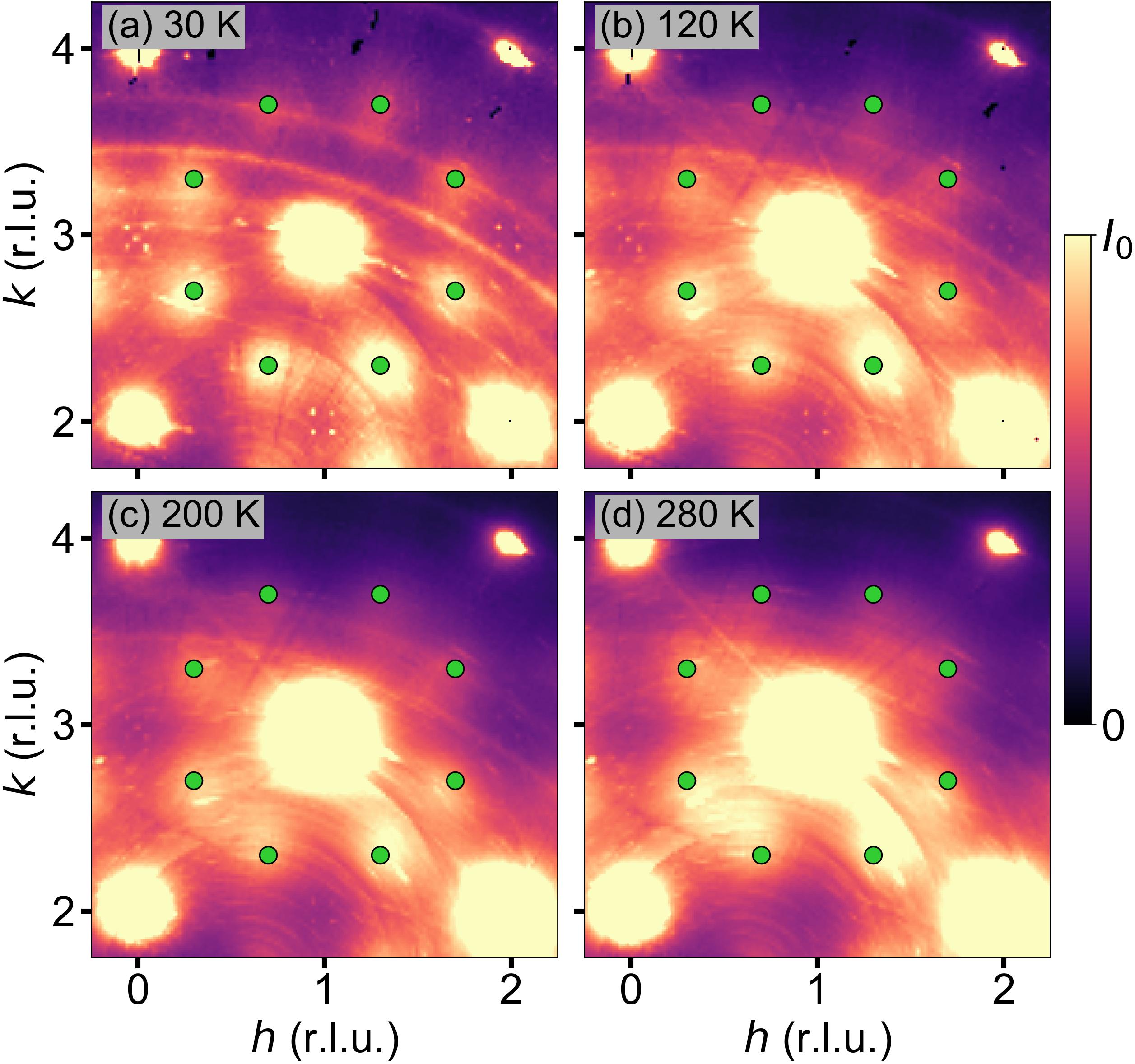} \\
\caption{\label{fig:octagon_satellites_move}
Figures showing the variation in position with temperature of the HK4 plane octagon satellites around the (134) main reflection. (a) Shows the reciprocal lattice pattern at 30 K, (b) at 120 K, (c) at 200 K, and (d) at 280 K. Green circles are used to mark the approximate locations of the satellites in the 30 K data. The circles remain unmoved from the 30 K positions in the higher temperature data to demonstrate the small shift in satellite position with temperature.
} 
\end{figure}

\begin{figure}
\centering\includegraphics[width=\columnwidth]{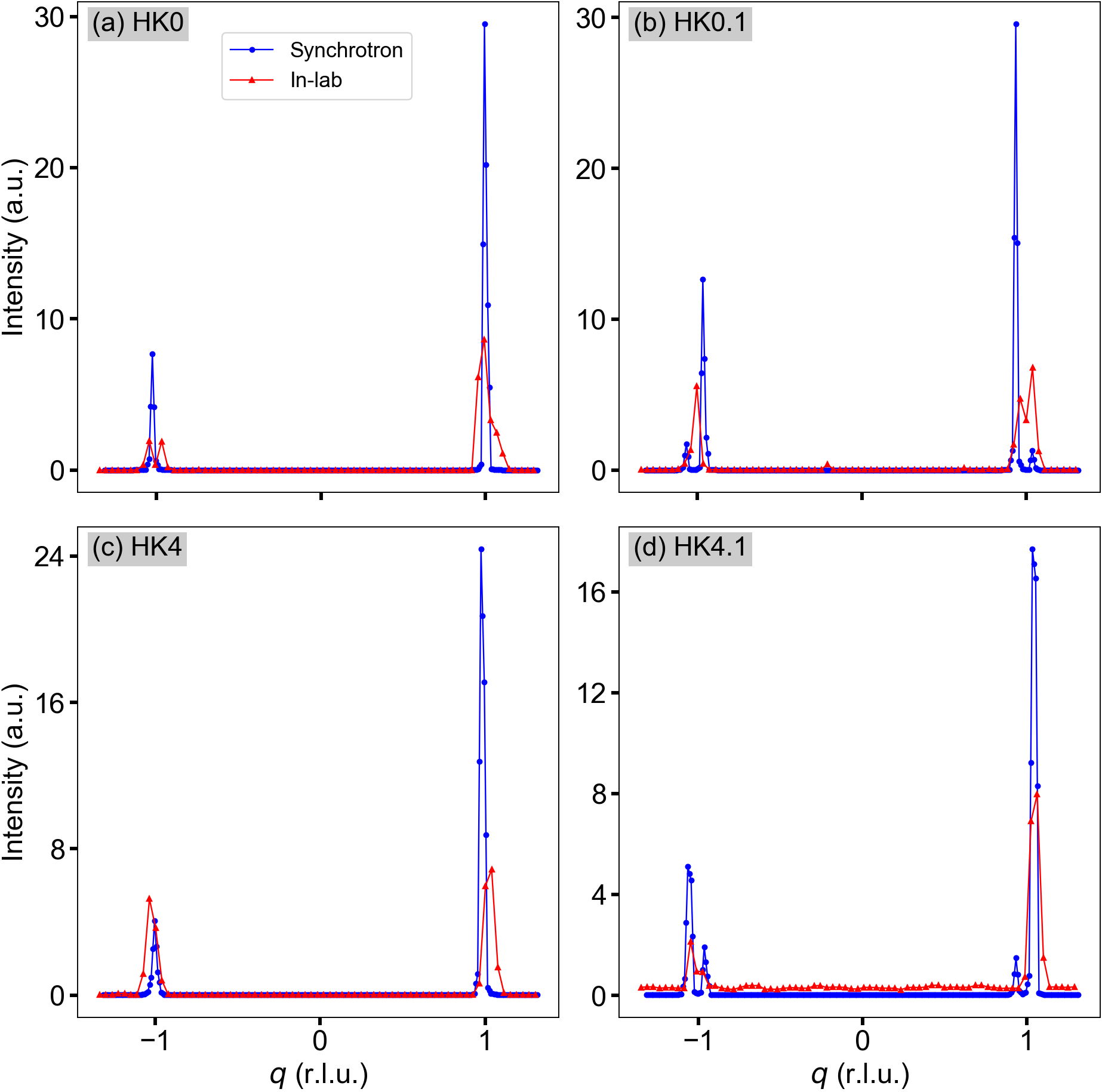} \\
\caption{\label{fig:TSI_CDW_Peak_Comparison}
Linecuts taken through both Bragg peaks and CDW satellite peaks for a comparison between the synchrotron and in-lab X-ray data. Synchrotron data is marked as blue circles, in-lab data as red triangles. (a, c) linecuts taken through two Bragg peaks found on the HK0 and HK4 planes, respectively. (b, d) linecuts taken through four CDW satellite peaks found on the HK plane at L = 0.1 and L = 4.1, respectively. The in-lab X-ray data shows a spillover of Bragg peak intensity into the space where the CDW satellites would be found, contributing to the inability to resolve these peaks in the in-lab data. All data is normalized to the background subtracted area.}
\end{figure}

\begin{figure}
\centering\includegraphics[width=0.8\columnwidth]{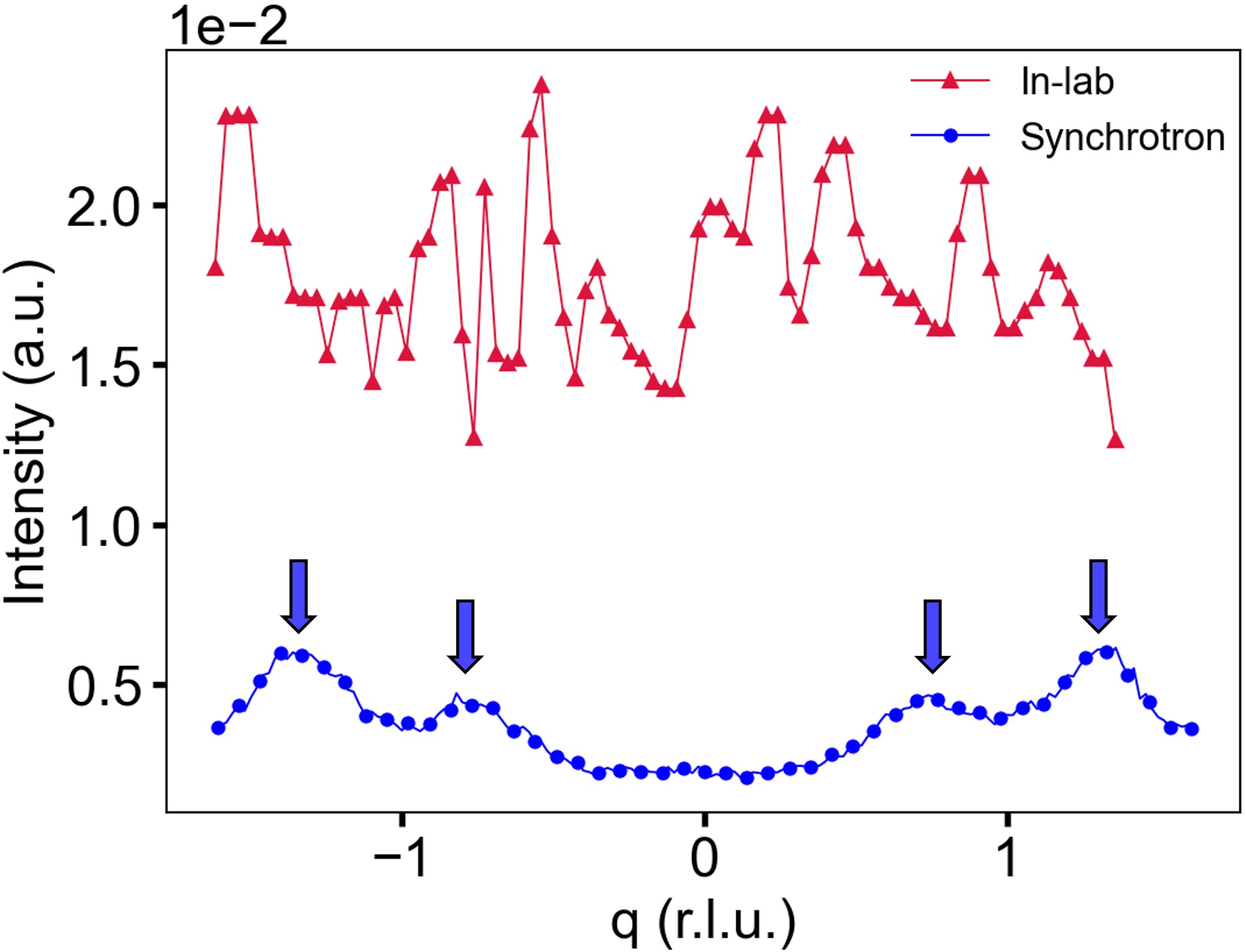} \\
\caption{\label{fig:Tsi_Toby_Experimental_Octagon_L=4_K=1.72}
Linecuts taken along the vector $[\textrm{H},1.72,4]$ through four diffuse octagon peaks. Synchrotron data is marked with blue circles, in-lab data is marked with red triangles. Data was collected at $T = 100$ K. Blue circles are plotted for every seven data points due to the high resolution of the synchrotron data. The curves are normalized using the background-subtracted integrated intensity of a selected Bragg peak from each set of data. The background of the in-lab data exceeds the intensity of the octagon peaks, an explanation for why these peaks are not observed in the in-lab data.
} 
\end{figure}


\end{center}

\clearpage
\printbibliography